\documentclass[aps,prd,twocolumn,superscriptaddress]{revtex4-2}

\usepackage{graphicx}
\usepackage{amsmath}
\usepackage{amssymb}
\usepackage[table]{xcolor}
\usepackage[colorlinks,citecolor=blue,linkcolor=blue,urlcolor=blue,anchorcolor=blue]{hyperref}

\allowdisplaybreaks

\newcommand\dx{{\rm d}}
\newcommand\p{\partial}
\newcommand\F{\mathcal{F}}
\newcommand\I{\mathcal{I}}
\newcommand\J{\mathcal{J}}
\newcommand\K{\mathcal{K}}
\newcommand\Q{\mathcal{Q}}
\newcommand\Y{\mathcal{Y}}
\newcommand\mond{\textsc{mond}}
\newcommand\GN{G_\textsc{n}}
\newcommand\KB{K_\textsc{b}}
\newcommand\SZ{{Skordis-Z\l{}o\'{s}nik }}

\newcommand{\pd}{\partial}

\newcommand\de{\textsc{de}}
\newcommand\f{\textsc{f}}

\newcommand\AaA{Astron. Astrophys.}
\newcommand\ApJ{Astrophys. J.}
\newcommand\ApJL{Astrophys. J. Lett.}

\newcommand\ARAA{Annu. Rev. Astron. Astrophys.}

\newcommand\JCAP{J. Cosmol. Astropart. Phys.}
\newcommand\LRR{Living Rev. Relativity}
\newcommand\MNRAS{Mon. Not. R. Astron. Soc.}
\newcommand\PLB{Phys. Lett. B}

\newcommand\PRD{Phys. Rev. D}
\newcommand\PRL{Phys. Rev. Lett.}

\newcommand\RMP{Rev. Mod. Phys.}

\newcommand\AnnPhysNY{Ann. Phys. (N.Y.)}

\newcommand\NatPhys{Nat. Phys.}

\newcommand\PhysDarkUniverse{Phys. Dark Universe}
\newcommand\PhysLettA{Phys. Lett. A}

\newcommand\PhysRev{Phys. Rev.}

\begin{document}

\title{Time evolution of the local gravitational parameters and gravitational wave polarizations in a relativistic MOND theory}
\author{Shuxun Tian}
\email[]{tshuxun@bnu.edu.cn}
\affiliation{Institute for Frontiers in Astronomy and Astrophysics, Beijing Normal University, Beijing 102206, China}
\affiliation{Department of Astronomy, Beijing Normal University, Beijing 100875, China}
\author{Shaoqi Hou}
\email[]{hou.shaoqi@whu.edu.cn}
\affiliation{School of Physics and Technology, Wuhan University, Wuhan 430072, China}
\author{Shuo Cao}
\email[]{caoshuo@bnu.edu.cn}
\affiliation{Institute for Frontiers in Astronomy and Astrophysics, Beijing Normal University, Beijing 102206, China}
\affiliation{Department of Astronomy, Beijing Normal University, Beijing 100875, China}
\author{Zong-Hong Zhu}
\email[]{zhuzh@bnu.edu.cn}
\affiliation{Institute for Frontiers in Astronomy and Astrophysics, Beijing Normal University, Beijing 102206, China}
\affiliation{Department of Astronomy, Beijing Normal University, Beijing 100875, China}
\affiliation{School of Physics and Technology, Wuhan University, Wuhan 430072, China}
\date{\today}
\begin{abstract}
  The recently proposed \SZ theory is the first relativistic MOND theory that can recover the success of the standard $\Lambda$CDM model at matching observations of the cosmic microwave background. This paper aims to revisit the Newtonian and MOND approximations and the gravitational wave analysis of the theory. For the local gravitational parameters, we show that one could obtain both time-varying effective Newtonian gravitational \textit{constant} $\GN$ and time-varying characteristic MOND acceleration scale $a_\mond$, by relaxing the static assumption extensively adopted in the literature. Specially, we successfully demonstrate how to reproduce the redshift dependence of $a_\mond$ observed in the \textit{Magneticum} cold dark matter simulations. For the gravitational waves, we show that there are only two tensor polarizations, and reconfirm that its speed is equal to the speed of light.
\end{abstract}
\maketitle

\section{Introduction}\label{sec:01}
Modified Newtonian dynamics (MOND) is an alternative to the dark matter paradigm, through the modification of Newton's law of universal gravitation or Newton's second law of motion \cite{Milgrom1983.ApJ.270.365,Milgrom1983.ApJ.270.371,Milgrom1983.ApJ.270.384}. The former belongs to the traditional modified gravity, and construction of its relativistic counterpart has been extensively discussed. \citet{Bekenstein1984.ApJ.286.7} proposed the first one. However, there are two major problems: the acausal problem \cite{Bekenstein1984.ApJ.286.7} and the gravitational lensing problem \cite{Bekenstein1994.ApJ.429.480}. Further modifications of the theory have been proposed to address these issues, such as the phase coupling \cite{Bekenstein1988.PLB.202.497} and disformal transformations \cite{Bekenstein1992.inproceedings,Bekenstein1993.PRD.48.3641,Sanders1997.ApJ.480.492,Bekenstein2004.PRD.70.083509}. These attempts made great progress in shaping the relativistic MOND theory and explaining the local gravitational phenomena \cite{Famaey2012.LRR.15.10}. However, for the cosmological linear perturbations, no such theory has been shown to successfully fit all the current data about the cosmic microwave background anisotropies and matter power spectra \cite{Skordis2006.PRL.96.011301,Dodelson2006.PRL.97.231301,Zuntz2010.PRD.81.104015,Xu2015.PRD.92.083505,Tan2018.JCAP.05.037}. Recently, \citet{Skordis2021.PRL.127.161302} proposed a new MOND theory to address this observational fitting problem. Analysis of this theory is the topic of this paper. In addition, we note that modification of Newton's second law still requires further development to arrive at a complete and observationally accepted theory \cite{Milgrom1994.AnnPhysNY.229.384,Milgrom1999.PhysLettA.253.273,Petersen2020.AaA.636.A56,Milgrom2022.PRD.106.064060}.

MOND theories generally predict an universal radial acceleration relation (RAR) --- correlation between the observed radial acceleration and that predicted by baryons with Newtonian gravity. \citet{McGaugh2016.PRL.117.201101} first observed the RAR in the SPARC database, and further data confirmed the conclusion \cite{Lelli2017.ApJ.836.152,Tian2020.ApJ.896.70}. This may be regarded as an observational evidence supporting MOND. However, after \citet{McGaugh2016.PRL.117.201101}, the same relation was also observed in the $N$-body simulations of cold dark matter (CDM) \cite{Dai2017.PRD.96.124016,Keller2017.ApJL.835.L17,Garaldi2018.PRL.120.261301,Dutton2019.MNRAS.485.1886}. The mass discrepancy acceleration relation, which is similar to RAR, was also predicted by MOND, and observed in both observations \cite{Durazo2017.ApJ.837.179} and CDM simulations \cite{Navarro2017.MNRAS.471.1841,Ludlow2017.PRL.118.161103}. In particular, \citet{Keller2017.ApJL.835.L17} found that the CDM simulated RAR depends on the cosmological redshift. This result implies that, in the framework of CDM, rotating galaxies still satisfy the \textit{universal} RAR at high redshifts. However, the parameter characterizing the acceleration scale in RAR is redshift-dependent. Recently, \citet{Mayer2023.MNRAS.518.257} presented an explicit redshift evolution of this characteristic MOND acceleration scale $a_\mond$ in the \textit{Magneticum} CDM simulations.

In the relativistic MOND theories, $a_\mond$ is a parameter and could be time-varying. \citet{Milgrom1983.ApJ.270.365} conjectured that $a_\mond\propto cH$ based on the numerical coincidence of their values at today. In the framework of T\textit{e}V\textit{e}S theory (a relativistic MOND theory) \cite{Bekenstein2004.PRD.70.083509}, \citet{Bekenstein2008.PRD.77.103512} analyzed this issue after considering the cosmological background evolution of the relevant fields. They found that $a_\mond$ changes on time scales much longer than the Hubble timescale. In this paper, we present the first analysis of the possible time evolution of the local Newtonian and MOND parameters in the \SZ theory \cite{Skordis2021.PRL.127.161302}. The method is principally the same as that in \cite{Bekenstein2008.PRD.77.103512}. Our result demonstrates how to reproduce the \textit{Magneticum} redshift dependence \cite{Mayer2023.MNRAS.518.257} in such relativistic MOND theory.

The first direct detection of the gravitational wave signal GW150914 has marked the new era of gravitational wave astronomy \cite{Abbott2016.PRL.116.061102}. In general relativity, there exist two well-known gravitational wave polarizations (plus and cross), traveling at the speed of
light. GW170814 and GW170817 observations confirmed these predictions \cite{Abbott2017.PRL.119.141101,Takeda2021.PRD.103.064037,Abbott2017.PRL.119.161101}. In this paper, we present a gauge-invariant gravitational wave analysis for the \SZ theory, in which the polarization content and the propagation speed are determined.

This paper is organized as follows. Section \ref{sec:02} introduces the \SZ MOND theory and summarizes the cosmic background evolutions. Note that, in principle, most of the results given in this section were obtained by \cite{Skordis2021.PRL.127.161302}. This section is retained to provide a clear basis for our subsequent discussions. Section \ref{sec:03} analyzes the Newtonian and MOND approximations. Section \ref{sec:04} discusses gravitational waves in the theory. Conclusions are presented in Sec. \ref{sec:05}.

Throughout this paper, we adopt the Hubble constant $H_0=67.4\,{\rm km}/{\rm s}/{\rm Mpc}$ and denote $h$ as its reduced value \cite{Aghanim2020.AaA.641.A6}. The subscript $0$ indicates the cosmological redshift $z=0$. In order to compare with observations, we adopt the SI Units and retain all physical constants in Sec. \ref{sec:02} and Sec. \ref{sec:03}. We set the speed of light $c=1$ in Sec. \ref{sec:04} for simplicity.

\section{The theory and cosmic evolutions}\label{sec:02}
The \SZ MOND theory is constructed based on a scalar field $\phi$ and a vector field $A_\mu$ \cite{Skordis2021.PRL.127.161302}. Its action is of the form $S = \int\dx^4x\sqrt{-g}\left[R + \mathcal{L}_\mond\right]/2\kappa + S_{\rm m}$, where $\kappa=8\pi\tilde{G}/c^4$ and $\tilde{G}$ is a constant with the same dimension of the Newtonian gravitational constant $\GN$. The MOND Lagrangian reads
\begin{align}
  & \mathcal{L}_\mond = - \frac{\KB}{2}F^{\mu\nu}F_{\mu\nu} + 2(2-\KB)J^\mu\nabla_\mu\phi \nonumber\\
  & \quad  - (2-\KB)\mathcal{Y} - \mathcal{F}(\mathcal{Y},\mathcal{Q})-\lambda(A^\mu A_\mu+1),
\end{align}
where $F_{\mu\nu}=2\nabla_{[\mu}A_{\nu]}$, $J_\mu=A^\alpha\nabla_\alpha A_\mu$, $\Y=q^{\mu\nu}\nabla_\mu\phi\nabla_\nu\phi$, $q^{\mu\nu}=g^{\mu\nu}+A^\mu A^\nu$, $\Q=A^\mu\nabla_\mu\phi$, $\F(\Y,\Q)$ is an arbitrary function, $\lambda$ is the Lagrange multiplier (a scalar), $\KB$ is a dimensionless constant. In our conventions, the dimension of $A_\mu$ relates to the metric ($[A^\mu A_\mu]=1$), $\phi$ is dimensionless, and $[\Y]=[\Q^2]=[\F]=[\lambda]={\rm length}^{-2}$.

The field equations can be derived from the variational principle. Variation of the action with respect to the metric gives the gravitational field equations
\begin{subequations}\label{eq:02}
\begin{align}\label{eq:02a}
  & G_{\mu\nu}  - \KB F_{\mu}^{\ \alpha}F_{\nu\alpha} + (2-\KB)\left\{ 2J_{(\mu}\nabla_{\nu)}\phi -A_\mu A_\nu\Box\phi \right. \nonumber\\
  & \ + \left. 2[A_{(\mu}\nabla_{\nu)} A_\alpha - A_{(\mu}\nabla_{|\alpha|} A_{\nu)}]\nabla^\alpha\phi  \right\} - \F_\Q A_{(\mu}\nabla_{\nu)}\phi \nonumber\\
  & \ - (2-\KB+\F_\Y)[\nabla_\mu\phi\nabla_\nu\phi+2\Q A_{(\mu}\nabla_{\nu)}\phi] \nonumber\\
  & \ - \lambda A_\mu A_\nu - g_{\mu\nu}\mathcal{L}_\mond/2 =\kappa T_{\mu\nu},
\end{align}
where $\F_\Y=\p\F/\p\Y$ and $\F_\Q=\p\F/\p\Q$. Variation of the action with respect to $\phi$ gives the scalar field equation
\begin{equation}\label{eq:02b}
  \nabla_\mu\mathcal{I}^\mu=0,
\end{equation}
where $\I^\mu=(2-\KB)J^\mu - (2-\KB+\F_\Y) q^{\alpha\mu}\nabla_\alpha\phi - \F_\Q A^\mu/2$. Variation of the action with respect to $A_\mu$ gives the vector field equations
\begin{align}\label{eq:02c}
  & \KB\nabla_\nu F^{\nu\mu} + (2-\KB)[(\nabla^\mu A_\nu)\nabla^\nu\phi - \nabla_\nu(A^\nu\nabla^\mu\phi)] \nonumber\\
  & \quad -\left[(2-\KB+\F_\Y)\Q + \F_\Q/2\right]\nabla^\mu\phi -\lambda A^\mu = 0.
\end{align}
Variation of the action with respect to $\lambda$ gives a constraint equation for the vector field
\begin{equation}\label{eq:02d}
  A^\mu A_\mu+1=0.
\end{equation}
\end{subequations}
Energy and momentum conservation $\nabla_\nu T^{\mu\nu}=0$ can be directly derived from Eq.~(\ref{eq:02}).

As we discussed in Sec. \ref{sec:01}, one goal of this paper is to study the time evolution of the local gravitational parameters in the \SZ MOND theory. In a relativistic theory, parameters describing the local gravitational system can be time-varying due to the cosmic evolution of the relevant fields. For example, the effective Newtonian gravitational \textit{constant} is time-varying in scalar-tensor gravity \cite{Brans1961.PhysRev.124.925,Damour1990.PRL.64.123,Babichev2011.PRL.107.251102,Zhang2019.PRD.100.024038,Burrage2020.JCAP.07.060} and nonlocal gravity \cite{Barreira2014.JCAP.09.031,Belgacem2019.JCAP.02.035,Tian2019.PRD.99.064044}. Here we summarize the cosmic background evolution for the \SZ MOND theory. We assume the Universe is described by the flat Friedmann-Lema{\^i}tre-Robertson-Walker (FLRW) metric $\dx s^2=-c^2\dx t^2+a^2\dx\mathbf{x}^2$,
where $a=a(t)$. To be consistent with Eq.~(\ref{eq:02d}), we assume $A_\mu=[-c,0,0,0]$ for the vector field. The scalar field is assumed to be $\phi=\phi(t)$. For the normal matters, we adopt $T^\mu_{\ \nu}={\rm diag}\{-\rho_{\rm m}c^2,p_{\rm m},p_{\rm m},p_{\rm m}\}$ \cite{Dodelson2020.book}. Substituting Eq.~(\ref{eq:02c}) into Eq.~(\ref{eq:02a}) eliminates $\lambda$. Then, substituting the above assumptions into the result, we obtain
\begin{subequations}\label{eq:03}
\begin{gather}
  H^2=\frac{8\pi\tilde{G}}{3}\rho_{\rm m}+\frac{c^2}{6}(\mathcal{F}-\Q\F_\Q), \label{eq:03a}\\
  H^2 + 2\frac{\ddot{a}}{a}=-\frac{8\pi\tilde{G}}{c^2} p_{\rm m} + \frac{1}{2}\mathcal{F}c^2, \label{eq:03b}
\end{gather}
with the cosmic background values $\Y=0$ and $\Q=\dot{\phi}/c$. Here the Hubble parameter $H\equiv\dot{a}/a$, $\dot{}\equiv\dx/\dx t$, and $\F$~and $\F_\Q$ are evaluated at the background. Equation~(\ref{eq:02b}) gives
\begin{equation}\label{eq:03c}
  \frac{\dx\F_\Q}{\dx t}+3H\F_\Q=0.
\end{equation}
\end{subequations}
To test self-consistency, we confirm that Eq.~(\ref{eq:02c}) gives only trivial results except one constraint equation on $\lambda$. Energy conservation $\dot{\rho}_{\rm m}+3H(\rho_{\rm m}+p_{\rm m}/c^2)=0$ can be derived from Eq.~(\ref{eq:03}) for arbitrary $\F$ function. In other words, Eqs.~(\ref{eq:03a}), (\ref{eq:03c}) and the matter energy conservation equation form a complete and self-consistent set. Based on Eq.~(\ref{eq:03}), we can define the effective MOND (dark matter) mass density and pressure as
\begin{equation}
  \rho_\mond = \frac{c^2}{16\pi\tilde{G}}(\mathcal{F}-\Q\F_\Q), \
  p_\mond = -\frac{\F c^4}{16\pi\tilde{G}}.
\end{equation}
Then Eq.~(\ref{eq:03}) can be rewritten as the two conventional Friedmann equations and one effective MOND energy conservation equation. The MOND relative mass density is defined as $\Omega_\mond\equiv8\pi\tilde{G}\rho_\mond/(3H^2)$.

\begin{figure}[!t]
  \centering
  \includegraphics[width=0.95\linewidth]{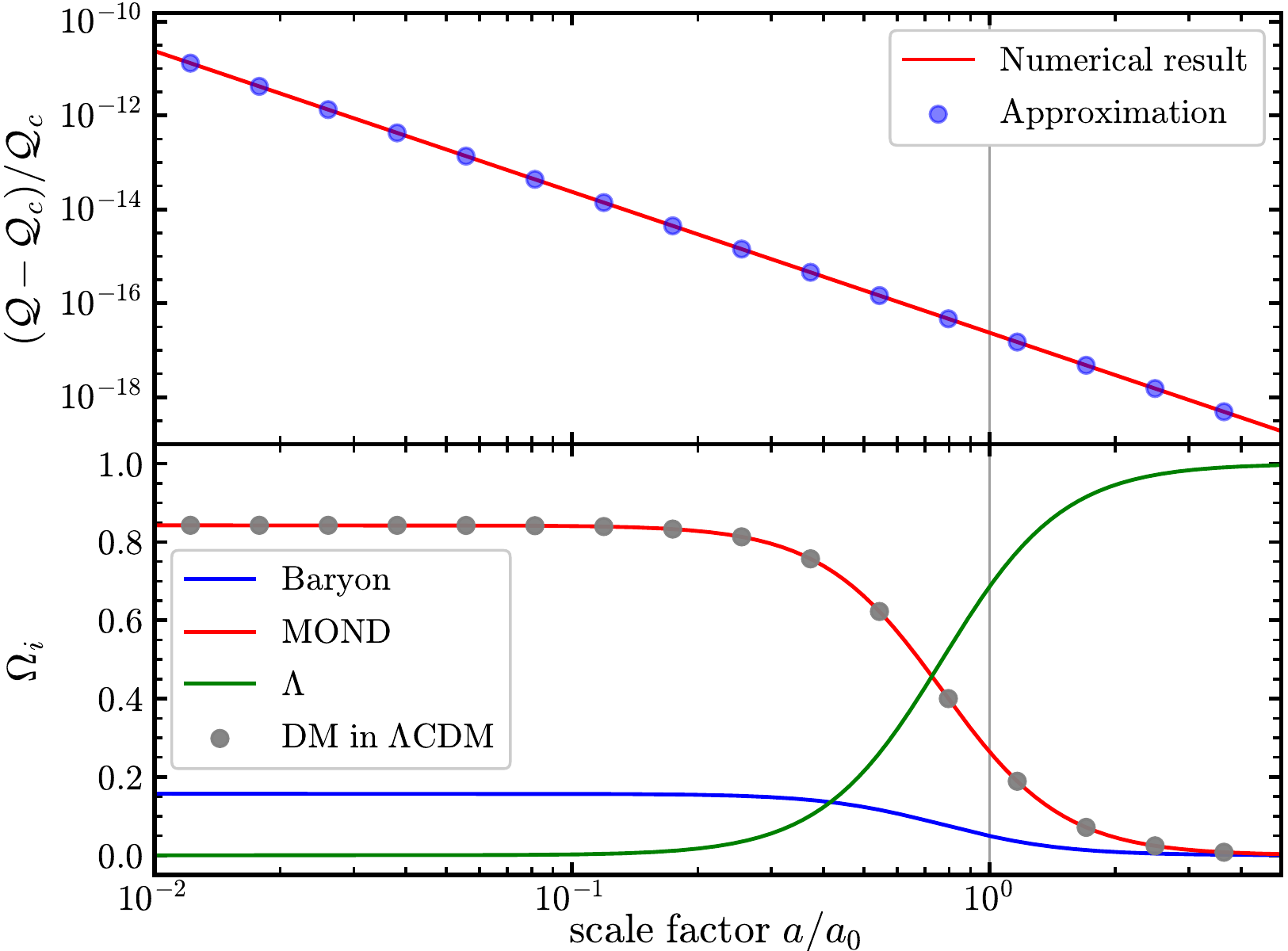}
  \caption{The cosmic background evolutions for the \SZ MOND theory (see the main text).}
  \label{fig:01}
\end{figure}

In order to reveal the key properties of the cosmic background evolution, and to be consistent with the conventions adopted in \cite{Skordis2021.PRL.127.161302}, we rewrite the function
\begin{equation}
  \F(\Y,\Q) = (2-\KB)\J(\Y,\Q) - 2\K(\Q).
\end{equation}
Here the first term satisfies $\J(0,\Q)=0$, and is used to produce the MOND behavior (see Sec. \ref{sec:03}). For the second term, we adopt the Higgs-like function $\K(\Q)=(\K_2/4\Q_c^2)(\Q^2-\Q_c^2)^2$, where $\K_2$ and $\Q_c$ are constant parameters \cite{Skordis2021.PRL.127.161302,Note1}.
The $\Q$ approaches to $\Q_c$ in the infinite future. In the late-time Universe, we can adopt $\rho_{\rm m}=\rho_{\rm baryon}\propto a^{-3}$. Meanwhile we add the cosmological constant $\Lambda$ to Eq.~(\ref{eq:03a}). Then Eq.~(\ref{eq:03c}) completely determines the cosmological evolution of $\Q$. The solid lines in Fig.~\ref{fig:01} show the numerical results for this ordinary differential system. The initial condition of $\Q$ and parameters of baryon and $\Lambda$ are set so that $\Omega_{\rm Baryon}h^2=0.0224$ and $\Omega_\mond h^2=0.120$ at today \cite{Aghanim2020.AaA.641.A6}. The MOND parameters are $\K_2=8.5\times10^8$ and $\Q_c=1\,{\rm Mpc}^{-1}$, which guarantee good fits to the \textit{Planck} CMB measurements and SDSS matter power spectra results \cite{Skordis2021.PRL.127.161302}. We set $\tilde{G}=\GN$, which corresponds to a special kind of $\J$ function (see Sec. \ref{sec:03}). Note that the cosmic background evolutions are independent of $\J$. We emphasize that high precision calculations are required to suppress numerical errors. The bottom part plots the relative energy density $\Omega_i$ of each component together with the result of dark matter in the standard $\Lambda$CDM model \cite{Aghanim2020.AaA.641.A6}. The coincidence between MOND and CDM indicates that such MOND can behave as cold as the CDM in the expanding Universe. The top part plots the evolution of $\Q$. Considering the extremely small value of the dimensionless $y$-axis, we conclude that, during the late-time era, $\Q$ keeps almost constant while the variation of $\Q-\Q_c$ is considerable. Generalizing the Taylor expansion discussed in \cite{Skordis2021.PRL.127.161302}, we obtain
\begin{subequations}\label{eq:06}
\begin{gather}
  \frac{\Q-\Q_c}{\Q_c} = \frac{1}{2}\beta-\frac{3}{8}\beta^2+\mathcal{O}(\beta^3), \label{eq:06a}\\
  \rho_\mond = \frac{\K_2c^2\Q_c^2}{8\pi\tilde{G}}\left[\beta+\frac{1}{4}\beta^2+\mathcal{O}(\beta^3)\right], \\
  p_\mond = \frac{\K_2c^4\Q_c^2}{32\pi\tilde{G}}\beta^2+\mathcal{O}(\beta^3), \\
  w_\mond = \frac{\beta}{4}+\mathcal{O}(\beta^2),
\end{gather}
where
\begin{equation}
  \beta=\frac{3\Omega_{\mond,0}H_0^2}{\K_2c^2\Q_c^2}\left(\frac{a_0}{a}\right)^3 \ll 1. \label{eq:06e}
\end{equation}
\end{subequations}
Note that $1+z=a_0/a$, where $z$ is the cosmological redshift. The $\Omega_{\mond,0}$ appearing in Eq.~(\ref{eq:06e}) can be regarded as a boundary condition of the differential equation (\ref{eq:03}). The above result confirms that the MOND is cold for the previous parameter settings. In the top part of Fig.~\ref{fig:01}, we also plot the leading term of Eq.~(\ref{eq:06a}), and the result shows it is a good approximation. Especially, the leading terms of Eq.~(\ref{eq:06}) are valid for a general $\K(\Q)$ once it satisfies $\K\approx\K_2(\Q-\Q_c)^2$ when $\Q\rightarrow\Q_c$.

\section{Newtonian and MOND analysis}\label{sec:03}
In the \SZ theory, the form of $\J(\Y,\Q)$ determines the local gravitational behaviors. \citet{Skordis2021.PRL.127.161302} pointed out two key properties. For physically acceptable scenarios, in the strong field region \cite{Note2},
the scalar field is described by the tracking or screening solution, which corresponds to the strong asymptotic expression $\J\rightarrow\Y$ or $\J\rightarrow\Y^p$ with $p\geq3/2$, respectively. The strong field solution determines the relation between $\tilde{G}$ and $\GN$. In the weak field region \cite{Note2}, MOND appears if $\J\rightarrow\Y^{3/2}$. These analyzes assumed that $\Q$ appearing in $\J$ reaches its cosmological minimum $\Q_c$. However, this may be invalid if variables such as $\Q-\Q_c$ appear in $\J$ (see Fig.~\ref{fig:01} for the evolutions). Similar to the time-varying $\GN$ in the scalar-tensor theory \cite{Brans1961.PhysRev.124.925,Damour1990.PRL.64.123,Babichev2011.PRL.107.251102,Zhang2019.PRD.100.024038,Burrage2020.JCAP.07.060}, relaxing this static assumption may make the local MOND parameters time-varying.

Considering the great success of the $\Lambda$CDM model in both theories and observations \cite{Ostriker1993.ARAA.31.689,Frieman2008.ARAA.46.385,Bertone2018.RMP.90.045002}, we wish to answer what kind of $\J(\Y,\Q)$ can reproduce the redshift dependence found in the \textit{Magneticum} $\Lambda$CDM simulations \cite{Mayer2023.MNRAS.518.257}. However, we emphasize that the \textit{Magneticum} trend has not been confirmed by observations. The difference between the MOND predictions and the \textit{Magneticum} trend does not mean the failure of either theory. Instead, this possible difference provides an indicator to distinguish between MOND and $\Lambda$CDM observationally in the future. Besides MOND, dark sector models beyond $\Lambda$CDM, such as dynamical dark energy and ultralight dark matter, might also predict a different $a_\mond$--$z$ relation. This is due to the fact that these models could affect galaxy formation \cite{Penzo2014.MNRAS.442.176,Schive2014.NatPhys.10.496}. A complete model dictionary of $a_\mond(z)$ is useful for future observational tests. The present paper only focuses on the part about the \SZ MOND theory.

Following \cite{Skordis2021.PRL.127.161302}, we adopt the perturbed metric $\dx s^2=-c^2(1+2\Phi/c^2)\dx t^2+(1-2\Phi/c^2)\dx\mathbf{x}^2$, where the first-order infinitesimal $\Phi=\Phi(\mathbf{x})$. The vector field is assumed to be $A_\mu=[-c(1+\Phi/c^2),0,0,0]$, which is consistent with Eq.~(\ref{eq:02d}). The scalar field is assumed to be $\phi=\bar{\phi}(t)+\varphi$, where the bar means cosmic background value and the first-order infinitesimal $\varphi=\varphi(\mathbf{x})$. The time derivative of the first-order infinitesimal is ignored because it is much smaller than the corresponding space derivative \cite{Skordis2021.PRL.127.161302,Tian2019.PRD.99.064044}. The possible time dependence of the local MOND parameters is encoded in $\bar{\phi}(t)$, or strictly $\bar{\Q}(t)$. Note that we no longer assume $\bar{\Q}=\Q_c$. Calculating the quadratic terms in the action with the above perturbations, we obtain
\begin{subequations}\label{eq:07}
\begin{align}\label{eq:07a}
  S^{(2)} &= -\int c\dx t\,\dx^3\mathbf{x}\left\{\frac{2-\KB}{16\pi\tilde{G}}\left[|\nabla\hat{\Phi}|^2 \right.\right. \nonumber\\
    &  \quad \left.\left. + c^4\J(\Y,\bar{\Q}) + \mathrm{mass\ terms}\right] + \rho\Phi\right\},
\end{align}
where
\begin{equation}
  \Y = |\nabla\varphi|^2+\mathrm{mass\ terms}, \label{eq:07b}
\end{equation}
\end{subequations}
and $\hat{\Phi}=\Phi-\varphi c^2$, $\rho$ is the local baryon mass density, and the mass terms indicate terms like ${\rm const.}\times\Phi^n$. The $\K(\Q)$ only contributes to the mass terms because $\Q=\bar{\Q}\cdot(1-\Phi/c^2+2\Phi^2/c^4)$. For the same reason, we can rewrite $\bar{\Q}$ as $\Q$ in Eq. (\ref{eq:07a}), and only consider the perturbation of $\Y$ in the following discussions. Hereafter we ignore the mass terms. This is reasonable because suitable parameters can indeed suppress the corresponding influences on the Newtonian and MOND dynamics \cite{Skordis2021.PRL.127.161302}. Integration by parts is used to eliminate the second derivative terms (e.g., $\Phi\nabla^2\Phi$) and obtain the above results.  Equation~(\ref{eq:07}) recovers Eq.~(6) in \cite{Skordis2021.PRL.127.161302} when $\bar{\Q}=\Q_c$. Hereafter we omit the bar in $\bar{\Q}$ and adopt $\J_\Y=\p\J(\Y,\Q)/\p\Y$. Variation of $S^{(2)}$ with respect to $\hat{\Phi}$ and $\varphi$, we obtain
\begin{subequations}\label{eq:08}
\begin{gather}
  \nabla^2\hat{\Phi}=\frac{8\pi\tilde{G}}{2-\KB}\rho, \\
  \nabla[\J_\Y\nabla\varphi]=\frac{8\pi\tilde{G}}{(2-\KB)c^2}\rho, \label{eq:08b}
\end{gather}
\end{subequations}
respectively. The \SZ theory is written in the Einstein frame with minimally coupling between matter and other fields. Therefore, $\Phi=\hat{\Phi}+\varphi c^2$ is the physical gravitational potential. In the weak field region, if $\J\propto\Y^{3/2}$, i.e., $\J_\Y\propto|\nabla\varphi|$, then $\varphi$ dominates $\Phi$ and produces the MOND behavior \cite{Bekenstein1984.ApJ.286.7,Skordis2021.PRL.127.161302}.

Here we discuss the possible time evolution of $\GN$ and $a_\mond$. Comparison of Eq.~(\ref{eq:08}) and Poisson equation in strong field region determines $\GN$. In the scaling case \cite{Skordis2021.PRL.127.161302}, we assume $\J\rightarrow\lambda_s\Y$, where the dimensionless variable $\lambda_s=\lambda_s(\Q)$. Then Eq.~(\ref{eq:08}) gives $\varphi c^2\rightarrow\hat{\Phi}/\lambda_s$ and
\begin{equation}\label{eq:09}
  \GN = \frac{2\tilde{G}}{2-\KB}(1+\frac{1}{\lambda_s}).
\end{equation}
Note that $\tilde{G}$ is a constant introduced in the action, and $\GN$ could be time-varying because of its dependence on $\lambda_s$. Considering $(\Q-\Q_c)/\Q_c\ll1$ (see Fig.~\ref{fig:01}), if $\lambda_s\propto\Q^p$, where $p$ is a constant, then the time evolution of $\GN$ is unobservable. However, if $\lambda_s\propto(\Q-\Q_c)^p\propto a^{-3p}$, then Eq.~(\ref{eq:09}) gives
\begin{equation}
  \frac{\dot{G}_\textsc{n}}{\GN} \approx -\frac{\dot{\lambda}_s}{\lambda_s^2}
  \approx\frac{3pH_0}{\lambda_{s,0}}.
\end{equation}
in which we assumed $\lambda_s\gg1$ and the last equality is valid at the low redshift Universe. Current observations give $|\dot{G}_\textsc{n}/\GN|\lesssim10^{-12}\,{\rm yr}^{-1}\approx0.01H_0$ \cite{Williams2004.PRL.93.261101,Hofmann2010.AaA.522.L5,Zhu2015.ApJ.809.41}. Therefore we require $\lambda_{s,0}\gtrsim100$ for this case. The screening case \cite{Skordis2021.PRL.127.161302} corresponds to $\lambda_s=\infty$, and results in an exactly constant $\GN$. This requires $\J\propto\Y^p$, where $p\geq3/2$ \cite{Skordis2021.PRL.127.161302}.

\begin{table*}[!t]
  \centering
  \tabcolsep=0cm
  \rowcolors{3}{lightgray!40}{}
  \begin{minipage}{0.77\textwidth}
  \caption{Models discussing the possible cosmological evolution of $a_\mond$. Note that Eqs. (\ref{eq:12}) and (\ref{eq:13}) are extensions of the original \SZ theory.}
  \label{tab:01}
  \begin{tabular}{@{\,}p{2.13cm}p{3.5cm}p{8.1cm}@{\,}}
    \hline
    \multicolumn{3}{c}{
    \footnotetext{Most of the existing relativistic MOND theories give constant $a_\mond$. Here we only consider the models that the constant $a_\mond$ can still be obtained after analyzing the relevant cosmic background evolutions.}
    \footnotetext{Here $l_\textsc{de}$ is the characteristic length scale of dark energy, which is of the order of $\Lambda^{-1/2}$, i.e., $c/H_0$, at today and could be time-varying in the dynamical models. }}
    \\[-11pt]
    \hline
    $a_\textsc{mond}(z)$                       & \SZ theory                                       & Other theories \& Phenomenological motivations \\
    \hline
    \vspace{-3.6pt}$\sim{\rm const.}$\footnotemark[1] & \vspace{-3.6pt}$\J=c_1\Y^{3/2}/\Q$        & T\textit{e}V\textit{e}S
                                                                                                    theory~\cite{Famaey2007.PRD.75.063002,Bekenstein2008.PRD.77.103512}
                                                                                                    $\qquad$ $\qquad$ $\qquad$ $\qquad$ $\qquad$ $\qquad$ $\quad$
                                                                                                    a subclass of nonlocal MOND models \cite{Deffayet2014.PRD.90.064038} \\
    $\propto(1+z)^3$                           & $\J=c_1\Y^{3/2}/(\Q-\Q_c)$                       &  --- \\
    \vspace{-3.6pt}$\propto cH(z)$             & \vspace{-3.6pt}$\J=c_1\Y^{3/2}/\nabla_\mu A^\mu$ & the numerical coincidence between $a_{\textsc{mond},0}$ and $cH_0$
                                                                                                    \cite{Milgrom1983.ApJ.270.365,Milgrom2015.PRD.91.044009}
                                                                                                    a subclass of nonlocal MOND models~\cite{Deffayet2014.PRD.90.064038} \\
    \vspace{-3.6pt}$\propto c^2/l_\textsc{de}$\footnotemark[2] & \vspace{-3.6pt}$\J=c_1\Y^{3/2}/\sqrt{V_\textsc{de}}$ & the numerical coincidence between $a_{\textsc{mond},0}$ and
                                                                                                    $c^2/l_{\textsc{de},0}$~\cite{Milgrom2015.PRD.91.044009}
                                                                                                    relativistic theories linking MOND to dark energy
                                                                                                    \cite{Zhao2007.ApJL.671.L1,Blanchet2008.PRD.78.024031} \\
    \textit{Magneticum}                        & Eq.~(\ref{eq:11})                                & --- \\
    \hline
    \hline
  \end{tabular}
  \end{minipage}
\end{table*}

\begin{figure}[!b]
  \centering
  \includegraphics[width=0.95\linewidth]{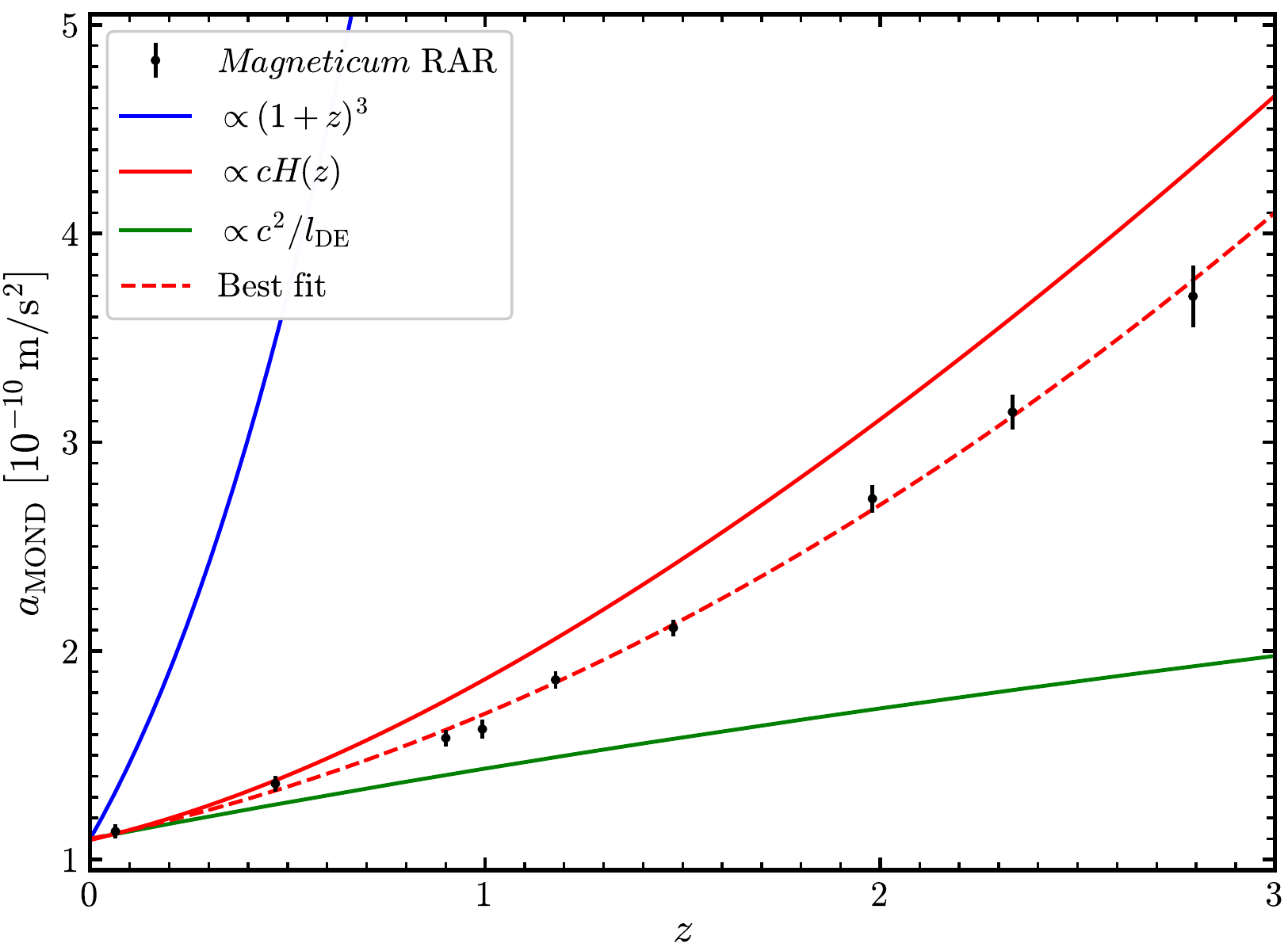}
  \caption{Four theoretical and \textit{Magneticum} simulated $a_\mond$ as a function of redshift. All results are calibrated at $z=0$.}
  \label{fig:02}
\end{figure}

The parameter $a_\mond$ is determined in the weak field region, in which $\varphi$ dominates $\Phi$. Hereafter, for simplicity, we adopt $\J\propto\Y^{3/2}$ throughout the MOND region to the Newtonian region. Considering Eqs.~(\ref{eq:07b}), (\ref{eq:08b}), (\ref{eq:09}) with $\lambda_s=\infty$, and Eq.~(3) in \cite{Bekenstein1984.ApJ.286.7}, we see that the coefficient of $\J\propto\Y^{3/2}$ equals to $2c^2/(3a_\mond)$. On the other hand, the coefficient can be written as a function of $\Q$. Theoretically, $a_\mond$ could be redshift-dependent. Here we discuss several explicit cases. Considering the dimensions of the variables, one of the simplest cases is $\J=c_1\Y^{3/2}/\Q$, where $c_i$ ($i=1,2,3\cdots$) is dimensionless constants. This case gives nearly constant $a_\mond$ in the late-time Universe (see Fig. \ref{fig:01}). Replacing $\Q$ with $\Q-\Q_c$ in the denominator, we obtain $a_\mond\propto(1+z)^3$ based on Eq.~(\ref{eq:06a}). Figure \ref{fig:02} depicts this result together with the \textit{Magneticum} result \cite{Mayer2023.MNRAS.518.257}. We see that this simple case fails to accurately describe the \textit{Magneticum} trend. The best polynomial fit of the \textit{Magneticum} result is $a_\mond=[0.9+0.2(1+z)^2]\times10^{-10}\,{\rm m}/{\rm s}^2$ \cite{Note3}.
Therefore, in the \SZ MOND theory, the \textit{Magneticum} trend could be reproduced by
\begin{equation}\label{eq:11}
  \J(\Y,\Q)=\frac{\Y^{3/2}}{c_1\Q_c+c_2\Q_c^{1/3}(\Q-\Q_c)^{2/3}}.
\end{equation}
For the parameters used in Sec. \ref{sec:02}, we obtain $c_1=4.6\times10^{-5}$ and $c_2=1.3\times10^6$ for the best fit. Note that Eq.~(\ref{eq:11}) gives an exactly constant $\GN$. Furthermore, if $\J\propto\Y^{3/2}$, then its specific form does not affect the cosmological linear perturbation analysis of the theory. The reason is that an equation similar to Eq.~(\ref{eq:07b}) can be obtained in the case of expanding Universe.

Equation (\ref{eq:11}) could reproduce the \textit{Magneticum} trend, but may not be the most natural way --- the functional form and relevant parameters require slight fine-tuning. Figure \ref{fig:02} also plots the case of $a_\mond\propto cH$, which is pretty close to the \textit{Magneticum} result. Inspired by the numerical coincidence between $a_{\mond,0}$ and $cH_0$, \citet{Milgrom1983.ApJ.270.365} firstly conjectured this relation. A theory that $a_\mond$ is controlled by $cH$ multiplied by a $\mathcal{O}(1)$-valued redshift-dependent function seems more natural. The \SZ theory may not be able to realize $a_\mond\propto cH$. The reason is that the explicit time-dependent variable that exists here is $\Q-\Q_c\propto(1+z)^3$, rather than an explicit $H$-dependent expression. Luckily, a minor extension of the \SZ theory can achieve the desired scenario. Considering $\nabla_\mu A^\mu=3H/c$ \cite{Deffayet2014.PRD.90.064038} (see also the Acknowledgements), we see that replacing $\J(\Y,\Q)$ with
\begin{equation}\label{eq:12}
  \J(\Y,\nabla_\mu A^\mu)=c_1\Y^{3/2}/\nabla_\mu A^\mu
\end{equation}
gives $a_\mond=2cH/c_1$, where the dimensionless constant $c_1\approx4\pi$ \cite{Milgrom2015.PRD.91.044009}. Note that, in the MOND analysis, we only need to consider the background value of $\nabla_\mu A^\mu$ because $\Y^{3/2}$ is a high-order infinitesimal. In addition, this extension does not destroy the success of the original \SZ theory in fitting the cosmological observations \cite{Skordis2021.PRL.127.161302}. This is due to the facts that $\Y=0$ in the cosmological background and $\Y^{3/2}$ only contributes the high-order terms in the cosmological linear perturbation analysis. For the same reason, in the framework of the \SZ theory with minor extension, we can link MOND to dark energy with
\begin{equation}\label{eq:13}
  \J=c_1\Y^{3/2}/\sqrt{V_\textsc{de}},
\end{equation}
where $V_\textsc{de}$ is the field potential of dark energy. Following the conventions in \cite{Tian2020.PRD.101.063531}, we know $c_1$ is dimensionless and $[V_\textsc{de}]={\rm length}^{-2}$. There are some work in the literature discussing possible links between MOND and dark energy \cite{Milgrom2015.PRD.91.044009,Zhao2007.ApJL.671.L1,Blanchet2008.PRD.78.024031}. Here Eq. (\ref{eq:13}) provides a new example. If dark energy is the cosmological constant, then this case gives constant $a_\mond$. However, if dark energy is dynamical, then $a_\mond$ could be redshift-dependent. In Fig.~\ref{fig:02}, the green line plots an illustration for the power-law potential with the index $p=1$ \cite{Peebles1988.ApJL.325.L17,Steinhardt1999.PRD.59.123504}. Detailed evolution of dark energy can be found in Appendix \ref{app:A}. We emphasize that current observations require $p<0.06$ \cite{Xu2022.PhysDarkUniverse.36.101023}, which in turn gives a flatter $a_\mond(z)$. Therefore, relativistic theory linking MOND to dark energy is not a good option to reproduce the \textit{Magneticum} trend.

Table \ref{tab:01} summarizes the models mentioned above. One thing is worth mentioning here. Generally, in the T\textit{e}V\textit{e}S theory \cite{Bekenstein2004.PRD.70.083509}, the $a_\mond$ changes much more slowly than $cH$ \cite{Bekenstein2008.PRD.77.103512}. To our knowledge, no specific T\textit{e}V\textit{e}S theory has been confirmed that can realize $a_\mond\propto cH$.

\section{Gravitational wave analysis}\label{sec:04}
Gravitational wave properties in the \SZ theory can be determined by solving the linearized equations of motion about the Minkowski spacetime defined by $\bar g_{\mu\nu}=\eta_{\mu\nu}$, $\bar A^\mu=\delta^\mu_0$, and $\bar\phi={\rm const.}$. This background solution requires that $\bar{\mathcal F}=\bar{\mathcal F}(\bar g_{\mu\nu}, \bar A^\mu,\bar\phi)=0$. The perturbed solutions are $g_{\mu\nu}=\eta_{\mu\nu}+h_{\mu\nu}$, $A^\mu=\delta^\mu_0+a^\mu$, and $\phi=\bar\phi+\varphi$. Since the linearized equations of motion are very complicated and coupled together, it is easier to use the gauge-invariant formalism to decouple the equations \cite{Flanagan:2005yc,Gong:2018cgj}. Following \citet{Gong:2018cgj}, one can decompose the components of $h_{\mu\nu}$ and $a^\mu$ as
\begin{subequations}
\begin{gather}
  h_{tt} = 2\phi, \label{httdec}\\
  h_{tj} = \beta_j+\partial_j\gamma, \label{htjdec}\\
  h_{jk} = h_{jk}^\mathrm{TT}+\frac{1}{3}H\delta_{jk}+\partial_{(j}\epsilon_{k)}+\left(\partial_j\partial_k-\frac{1}{3}\delta_{jk}\nabla^2\right)\rho,\label{hjkdec}\\
  a^t=\frac{1}{2}h_{tt}=\phi,\label{v0dec}\\
   a^j=\mu^j+\partial^j\omega.\label{vjdec}
\end{gather}
\end{subequations}
Here, $\pd^kh_{jk}^\mathrm{TT}=0$, $\eta^{jk}h_{jk}^\mathrm{TT}=0$, and $\pd_j\beta^j=\pd_j\epsilon^j=\pd_j\mu^j=0$.
Under the infinitesimal coordinate transformation parameterized by $\xi^\mu=(\xi^t,\xi^j)=(A,B^j+\pd^jC)$ with $\pd_jB^j=0$, one knows that
\begin{subequations}
\begin{gather}\label{eq-gt}
  h_{\mu\nu}\rightarrow h_{\mu\nu}-\partial_\mu\xi_\nu-\partial_\nu\xi_\mu,\\
  a^\mu\rightarrow a^\mu+\bar A^\nu\partial_\nu\xi^\mu,\\
  \varphi\rightarrow\varphi.
\end{gather}
\end{subequations}
Therefore, one determines the following gauge-invariant variables,
\begin{subequations}
\begin{gather}
    \varphi,\quad h^\mathrm{TT}_{jk},\\
  \Phi = -\phi+\dot\gamma-\frac{1}{2}\ddot\rho, \\
  \Theta = \frac{1}{3}(H-\nabla^2\rho), \\
  \Xi_j = \beta_j-\frac{1}{2}\dot\epsilon_j,\\
  \Sigma_j=\beta_j+\mu_j,\\
   \Omega=\omega+\frac{1}{2}\dot\rho.
\end{gather}
\end{subequations}
Then, one can try to reexpress the linearized equations of motion to conclude that
\begin{subequations}
\begin{gather}
    \ddot h_{jk}^\mathrm{TT}-\nabla^2h_{jk}^\mathrm{TT}=0,\\
    \ddot \Sigma_j-\nabla^2\Sigma_j=0,\\
    \frac{\pd^2\bar{\mathcal F}}{\pd\mathcal Q^2}\ddot\varphi+2\left[ \frac{2(2-\KB)}{\KB}+\frac{\pd\bar{\mathcal F}}{\pd\mathcal Y} \right]\nabla^2\varphi=0,\\
    \Xi_j=0,\\
    \Theta=\Phi=0,\\
    \dot\Omega=\frac{\KB-2}{\KB}\varphi,
\end{gather}
\end{subequations}
assuming $\pd\bar{\mathcal F}/\pd\mathcal Q=0$, where barred quantities are to be evaluated at the flat spacetime background. The above equations show that the tensor and vector modes are propagating at the speed of light, while $\varphi$ generally travels at a different speed, which is smaller than the speed of light for the parameter values adopted in Sec. \ref{sec:02}. This result reconfirmed the conclusion presented in \cite{Skordis2021.PRL.127.161302}.

Provided that the ordinary matter couples with the metric minimally, one can calculate the geodesic deviation equation, $\ddot x^j=-R_{tjtk}x^k$, to determine the polarizations of gravitational waves \cite{Misner:1974qy}. It turns of that
\begin{equation}
    \label{eq-relc}
    R_{tjtk}=-\frac{1}{2}h_{jk}^\mathrm{TT}.
\end{equation}
Therefore, there are only two polarizations (plus and cross), like in general relativity.

\section{Conclusions}\label{sec:05}
In this paper, we discuss the Newtonian, MOND and gravitational wave analyses for the \SZ theory \cite{Skordis2021.PRL.127.161302}. In the first two cases, after abandoning the static assumption adopted in \cite{Skordis2021.PRL.127.161302}, we find that whether $\GN$ and $a_\mond$ are time-varying depends on the specific form of the $\J$ function of the theory. Screening the scalar field in strong field region is a sufficient condition to give a constant $\GN$, and this scenario may be a preferred choice in both theory and observations. For the $a_\mond$, we highlight that the theory with Eq.~(\ref{eq:11}) could reproduce the $a_\mond$--$z$ dependence observed in the \textit{Magneticum} simulations \cite{Mayer2023.MNRAS.518.257}. Minor extension of the original \SZ theory with Eq. (\ref{eq:12}) gives $a_\mond\propto cH$. For the gravitational wave analysis, we show that there are only two tensor polarizations, which is preferred by the GW170814 observations \cite{Abbott2017.PRL.119.141101,Takeda2021.PRD.103.064037}.

\section*{Acknowledgements}
We especially thank the referee for pointing out $\nabla_\mu A^\mu\propto H(z)$ and suggesting that we discuss the model described by Eq. (\ref{eq:12}). This work was supported by the National Natural Science Foundation of China under Grants No. 11633001, No. 11920101003, No. 12021003 and No. 11690023, and the Strategic Priority Research Program of the Chinese Academy of Sciences, Grant No. XDB23000000. S. T. was supported by the Initiative Postdocs Supporting Program under Grant No. BX20200065 and China Postdoctoral Science Foundation under Grant No. 2021M700481. S. H. was supported by the National Natural Science Foundation of China under Grant No. 12205222.

\appendix
\section{Cosmic evolution of dark energy}\label{app:A}
\begin{figure}[!t]
  \centering
  \includegraphics[width=0.95\linewidth]{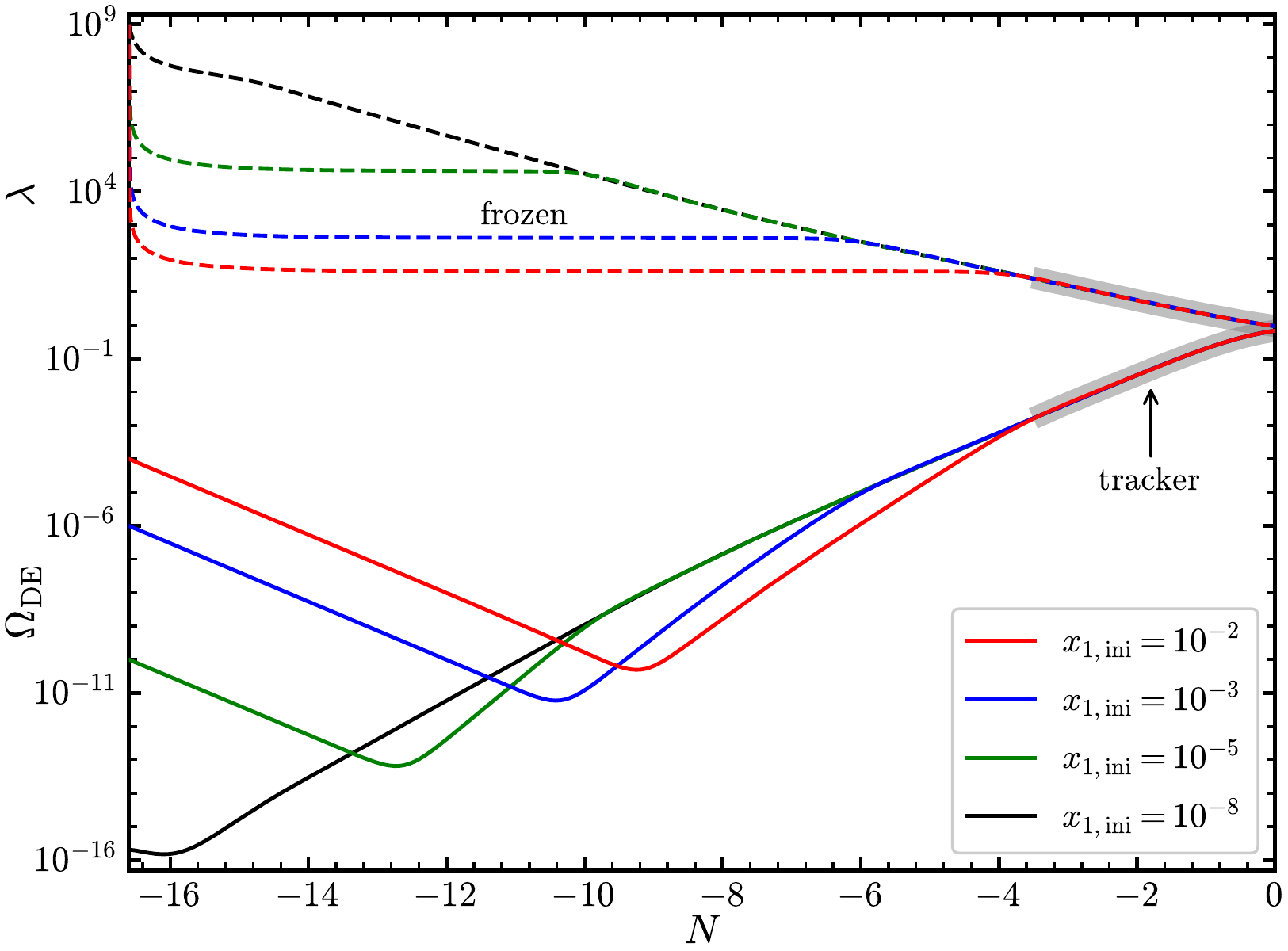}
  \caption{Cosmic evolution of the $\Omega_\de$ and $\lambda$ with model parameter $p=1$. Other parameters including $V_{\de,0}$ and $\phi_{\de,0}$ are not necessary to do this calculation. The initial conditions are $x_{1,{\rm ini}}=\{10^{-2},10^{-3},10^{-5},10^{-8}\}$, $x_{2,{\rm ini}}=10^{-8}$, $\lambda_{\rm ini}=10^9$ and $N_{\rm ini}=-16.60$. The above settings give $\Omega_{\de,0}\approx70\%$.}
  \label{fig:03}
\end{figure}

Equation (\ref{eq:13}) describes a model linking MOND to dark energy. We adopt a quintessence model with field potential $V_\de(\phi_\de)=V_{\de,0}\cdot(\phi_{\de,0}/\phi_\de)^p$ \cite{Peebles1988.ApJL.325.L17,Steinhardt1999.PRD.59.123504}, where the index $p\geq0$. Following the conventions in \cite{Tian2020.PRD.101.063531}, we have $[V_\de]={\rm length}^{-2}$. For the flat FLRW Universe, the cosmic evolution equations are \cite{Tian2020.PRD.101.063531}
\begin{subequations}
\begin{gather}
  H^2 = \frac{8\pi\tilde{G}}{3}(\rho_\f+\rho_\de), \\
  \ddot{\phi}_\de+3H\dot{\phi}_\de+c^2V_\de'=0, \\
  \dot{\rho}_\f+3(1+w_\f)H\rho_\f=0,
\end{gather}
\end{subequations}
where $'\equiv\dx/\dx\phi_\de$, $\rho_\de=(c^2/8\pi\tilde{G})\cdot[\dot{\phi}_\de^2/(2c^2)+V_\de]$ and the subscript $\textsc{f}$ means fluid. Here we regard MOND as a pressureless dark matter and include its contribution in $\rho_\f$. This is a good approximation as shown in Fig. \ref{fig:01}. Then the equation of state $w_\f$ is given by Eq. (4) in \cite{Tian2020.PRD.102.063509}. Introducing the dimensionless variables
\begin{gather}
  x_1=\frac{\dot{\phi}_\de}{\sqrt{6}H}, \quad
  x_2=\frac{c\sqrt{V_\de}}{\sqrt{3}H}, \quad
  \lambda=-\frac{V_\de'}{V_\de}=\frac{p}{\phi_\de}, \nonumber\\
  \Gamma=\frac{V_\de''V_\de}{(V_\de')^2}=\frac{p+1}{p},
\end{gather}
the above evolution equations can be rewritten as
\begin{subequations}\label{eq:A3}
\begin{align}
  \frac{\dx x_1}{\dx N} &= -3x_1+\frac{\sqrt{6}}{2}\lambda x_2^2+\frac{3}{2}x_1L, \\
  \frac{\dx x_2}{\dx N} &= -\frac{\sqrt{6}}{2}\lambda x_1x_2+\frac{3}{2}x_2L, \\
  \frac{\dx\lambda}{\dx N} &= \sqrt{6}\lambda^2(1-\Gamma)x_1,
\end{align}
\end{subequations}
where $L=(1-w_\f)x_1^2+(1+w_\f)(1-x_2^2)$ and $N=\ln(a/a_0)$. The relative dark energy density $\Omega_\textsc{de}=x_1^2+x_2^2$. Figure \ref{fig:03} presents the numerical solutions of Eq. (\ref{eq:A3}), and illustrates the frozen and tracker properties of this model \cite{Steinhardt1999.PRD.59.123504}. We can use the tracker solution to calculate $a_\mond(z)$ in the low-redshift Universe. Especially, we have $a_\mond\propto\sqrt{V_\de}\propto\lambda^{p/2}$. Considering the calibration at $z=0$ in Fig. \ref{fig:02}, we can directly obtain $a_\mond(z)$ from the solution of $\lambda$ without the value of $V_{\de,0}$. Parameters adopted in Fig. \ref{fig:03} are used to plot the green line in Fig. \ref{fig:02}.

\begin{thebibliography}{71}%
\makeatletter
\providecommand \@ifxundefined [1]{%
 \@ifx{#1\undefined}
}%
\providecommand \@ifnum [1]{%
 \ifnum #1\expandafter \@firstoftwo
 \else \expandafter \@secondoftwo
 \fi
}%
\providecommand \@ifx [1]{%
 \ifx #1\expandafter \@firstoftwo
 \else \expandafter \@secondoftwo
 \fi
}%
\providecommand \natexlab [1]{#1}%
\providecommand \enquote  [1]{``#1''}%
\providecommand \bibnamefont  [1]{#1}%
\providecommand \bibfnamefont [1]{#1}%
\providecommand \citenamefont [1]{#1}%
\providecommand \href@noop [0]{\@secondoftwo}%
\providecommand \href [0]{\begingroup \@sanitize@url \@href}%
\providecommand \@href[1]{\@@startlink{#1}\@@href}%
\providecommand \@@href[1]{\endgroup#1\@@endlink}%
\providecommand \@sanitize@url [0]{\catcode `\\12\catcode `\$12\catcode
  `\&12\catcode `\#12\catcode `\^12\catcode `\_12\catcode `\%12\relax}%
\providecommand \@@startlink[1]{}%
\providecommand \@@endlink[0]{}%
\providecommand \url  [0]{\begingroup\@sanitize@url \@url }%
\providecommand \@url [1]{\endgroup\@href {#1}{\urlprefix }}%
\providecommand \urlprefix  [0]{URL }%
\providecommand \Eprint [0]{\href }%
\providecommand \doibase [0]{https://doi.org/}%
\providecommand \selectlanguage [0]{\@gobble}%
\providecommand \bibinfo  [0]{\@secondoftwo}%
\providecommand \bibfield  [0]{\@secondoftwo}%
\providecommand \translation [1]{[#1]}%
\providecommand \BibitemOpen [0]{}%
\providecommand \bibitemStop [0]{}%
\providecommand \bibitemNoStop [0]{.\EOS\space}%
\providecommand \EOS [0]{\spacefactor3000\relax}%
\providecommand \BibitemShut  [1]{\csname bibitem#1\endcsname}%
\let\auto@bib@innerbib\@empty
\bibitem [{\citenamefont
  {{Milgrom}}(1983{\natexlab{a}})}]{Milgrom1983.ApJ.270.365}%
  \BibitemOpen
  \bibfield  {author} {\bibinfo {author} {\bibfnamefont {M.}~\bibnamefont
  {{Milgrom}}},\ }\bibinfo {title} {{A modification of the Newtonian dynamics
  as a possible alternative to the hidden mass hypothesis}},\ \href
  {https://doi.org/10.1086/161130} {\bibfield  {journal} {\bibinfo  {journal}
  {\ApJ}\ }\textbf {\bibinfo {volume} {270}},\ \bibinfo {pages} {365} (\bibinfo
  {year} {1983}{\natexlab{a}})}\BibitemShut {NoStop}%
\bibitem [{\citenamefont
  {{Milgrom}}(1983{\natexlab{b}})}]{Milgrom1983.ApJ.270.371}%
  \BibitemOpen
  \bibfield  {author} {\bibinfo {author} {\bibfnamefont {M.}~\bibnamefont
  {{Milgrom}}},\ }\bibinfo {title} {{A modification of the Newtonian dynamics:
  Implications for galaxies}},\ \href {https://doi.org/10.1086/161131}
  {\bibfield  {journal} {\bibinfo  {journal} {\ApJ}\ }\textbf {\bibinfo
  {volume} {270}},\ \bibinfo {pages} {371} (\bibinfo {year}
  {1983}{\natexlab{b}})}\BibitemShut {NoStop}%
\bibitem [{\citenamefont
  {{Milgrom}}(1983{\natexlab{c}})}]{Milgrom1983.ApJ.270.384}%
  \BibitemOpen
  \bibfield  {author} {\bibinfo {author} {\bibfnamefont {M.}~\bibnamefont
  {{Milgrom}}},\ }\bibinfo {title} {{A modification of the Newtonian dynamics:
  Implications for galaxy systems}},\ \href {https://doi.org/10.1086/161132}
  {\bibfield  {journal} {\bibinfo  {journal} {\ApJ}\ }\textbf {\bibinfo
  {volume} {270}},\ \bibinfo {pages} {384} (\bibinfo {year}
  {1983}{\natexlab{c}})}\BibitemShut {NoStop}%
\bibitem [{\citenamefont {{Bekenstein}}\ and\ \citenamefont
  {{Milgrom}}(1984)}]{Bekenstein1984.ApJ.286.7}%
  \BibitemOpen
  \bibfield  {author} {\bibinfo {author} {\bibfnamefont {J.}~\bibnamefont
  {{Bekenstein}}}\ and\ \bibinfo {author} {\bibfnamefont {M.}~\bibnamefont
  {{Milgrom}}},\ }\bibinfo {title} {{Does the missing mass problem signal the
  breakdown of Newtonian gravity?}},\ \href {https://doi.org/10.1086/162570}
  {\bibfield  {journal} {\bibinfo  {journal} {\ApJ}\ }\textbf {\bibinfo
  {volume} {286}},\ \bibinfo {pages} {7} (\bibinfo {year} {1984})}\BibitemShut
  {NoStop}%
\bibitem [{\citenamefont {{Bekenstein}}\ and\ \citenamefont
  {{Sanders}}(1994)}]{Bekenstein1994.ApJ.429.480}%
  \BibitemOpen
  \bibfield  {author} {\bibinfo {author} {\bibfnamefont {J.~D.}\ \bibnamefont
  {{Bekenstein}}}\ and\ \bibinfo {author} {\bibfnamefont {R.~H.}\ \bibnamefont
  {{Sanders}}},\ }\bibinfo {title} {{Gravitional lenses and unconventional
  gravity theories}},\ \href {https://doi.org/10.1086/174337} {\bibfield
  {journal} {\bibinfo  {journal} {\ApJ}\ }\textbf {\bibinfo {volume} {429}},\
  \bibinfo {pages} {480} (\bibinfo {year} {1994})}\BibitemShut {NoStop}%
\bibitem [{\citenamefont {Bekenstein}(1988)}]{Bekenstein1988.PLB.202.497}%
  \BibitemOpen
  \bibfield  {author} {\bibinfo {author} {\bibfnamefont {J.~D.}\ \bibnamefont
  {Bekenstein}},\ }\bibinfo {title} {{Phase coupling gravitation: Symmetries
  and gauge fields}},\ \href {https://doi.org/10.1016/0370-2693(88)91851-5}
  {\bibfield  {journal} {\bibinfo  {journal} {\PLB}\ }\textbf {\bibinfo
  {volume} {202}},\ \bibinfo {pages} {497} (\bibinfo {year}
  {1988})}\BibitemShut {NoStop}%
\bibitem [{\citenamefont {Bekenstein}(1992)}]{Bekenstein1992.inproceedings}%
  \BibitemOpen
  \bibfield  {author} {\bibinfo {author} {\bibfnamefont {J.~D.}\ \bibnamefont
  {Bekenstein}},\ }\bibinfo {title} {{New gravitational theories as
  alternatives to dark matter}},\ in\ \href
  {https://doi.org/10.1142/9789814537643} {\emph {\bibinfo {booktitle}
  {Proceedings of the Sixth Marcel Grossmann Meeting on General Relativity}}},\
  \bibinfo {editor} {edited by\ \bibinfo {editor} {\bibfnamefont
  {H.}~\bibnamefont {Sato}}\ and\ \bibinfo {editor} {\bibfnamefont
  {T.}~\bibnamefont {Nakamura}}}\ (\bibinfo  {publisher} {World Scientific},\
  \bibinfo {address} {Singapore},\ \bibinfo {year} {1992})\ pp.\ \bibinfo
  {pages} {905--924}\BibitemShut {NoStop}%
\bibitem [{\citenamefont {Bekenstein}(1993)}]{Bekenstein1993.PRD.48.3641}%
  \BibitemOpen
  \bibfield  {author} {\bibinfo {author} {\bibfnamefont {J.~D.}\ \bibnamefont
  {Bekenstein}},\ }\bibinfo {title} {{Relation between physical and
  gravitational geometry}},\ \href {https://doi.org/10.1103/PhysRevD.48.3641}
  {\bibfield  {journal} {\bibinfo  {journal} {\PRD}\ }\textbf {\bibinfo
  {volume} {48}},\ \bibinfo {pages} {3641} (\bibinfo {year}
  {1993})}\BibitemShut {NoStop}%
\bibitem [{\citenamefont {{Sanders}}(1997)}]{Sanders1997.ApJ.480.492}%
  \BibitemOpen
  \bibfield  {author} {\bibinfo {author} {\bibfnamefont {R.~H.}\ \bibnamefont
  {{Sanders}}},\ }\bibinfo {title} {{A stratified framework for scalar-tensor
  theories of modified dynamics}},\ \href {https://doi.org/10.1086/303980}
  {\bibfield  {journal} {\bibinfo  {journal} {\ApJ}\ }\textbf {\bibinfo
  {volume} {480}},\ \bibinfo {pages} {492} (\bibinfo {year}
  {1997})}\BibitemShut {NoStop}%
\bibitem [{\citenamefont {Bekenstein}(2004)}]{Bekenstein2004.PRD.70.083509}%
  \BibitemOpen
  \bibfield  {author} {\bibinfo {author} {\bibfnamefont {J.~D.}\ \bibnamefont
  {Bekenstein}},\ }\bibinfo {title} {{Relativistic gravitation theory for the
  modified Newtonian dynamics paradigm}},\ \href
  {https://doi.org/10.1103/PhysRevD.70.083509} {\bibfield  {journal} {\bibinfo
  {journal} {\PRD}\ }\textbf {\bibinfo {volume} {70}},\ \bibinfo {pages}
  {083509} (\bibinfo {year} {2004})}\BibitemShut {NoStop}%
\bibitem [{\citenamefont {{Famaey}}\ and\ \citenamefont
  {{McGaugh}}(2012)}]{Famaey2012.LRR.15.10}%
  \BibitemOpen
  \bibfield  {author} {\bibinfo {author} {\bibfnamefont {B.}~\bibnamefont
  {{Famaey}}}\ and\ \bibinfo {author} {\bibfnamefont {S.~S.}\ \bibnamefont
  {{McGaugh}}},\ }\bibinfo {title} {{Modified Newtonian dynamics (MOND):
  Observational phenomenology and relativistic extensions}},\ \href
  {https://doi.org/10.12942/lrr-2012-10} {\bibfield  {journal} {\bibinfo
  {journal} {\LRR}\ }\textbf {\bibinfo {volume} {15}},\ \bibinfo {eid} {10}
  (\bibinfo {year} {2012})}\BibitemShut {NoStop}%
\bibitem [{\citenamefont {Skordis}\ \emph {et~al.}(2006)\citenamefont
  {Skordis}, \citenamefont {Mota}, \citenamefont {Ferreira},\ and\
  \citenamefont {B\oe{}hm}}]{Skordis2006.PRL.96.011301}%
  \BibitemOpen
  \bibfield  {author} {\bibinfo {author} {\bibfnamefont {C.}~\bibnamefont
  {Skordis}}, \bibinfo {author} {\bibfnamefont {D.~F.}\ \bibnamefont {Mota}},
  \bibinfo {author} {\bibfnamefont {P.~G.}\ \bibnamefont {Ferreira}},\ and\
  \bibinfo {author} {\bibfnamefont {C.}~\bibnamefont {B\oe{}hm}},\ }\bibinfo
  {title} {{Large Scale Structure in Bekenstein's Theory of Relativistic
  Modified Newtonian Dynamics}},\ \href
  {https://doi.org/10.1103/PhysRevLett.96.011301} {\bibfield  {journal}
  {\bibinfo  {journal} {\PRL}\ }\textbf {\bibinfo {volume} {96}},\ \bibinfo
  {pages} {011301} (\bibinfo {year} {2006})}\BibitemShut {NoStop}%
\bibitem [{\citenamefont {Dodelson}\ and\ \citenamefont
  {Liguori}(2006)}]{Dodelson2006.PRL.97.231301}%
  \BibitemOpen
  \bibfield  {author} {\bibinfo {author} {\bibfnamefont {S.}~\bibnamefont
  {Dodelson}}\ and\ \bibinfo {author} {\bibfnamefont {M.}~\bibnamefont
  {Liguori}},\ }\bibinfo {title} {{Can Cosmic Structure Form without Dark
  Matter?}},\ \href {https://doi.org/10.1103/PhysRevLett.97.231301} {\bibfield
  {journal} {\bibinfo  {journal} {\PRL}\ }\textbf {\bibinfo {volume} {97}},\
  \bibinfo {pages} {231301} (\bibinfo {year} {2006})}\BibitemShut {NoStop}%
\bibitem [{\citenamefont {Zuntz}\ \emph {et~al.}(2010)\citenamefont {Zuntz},
  \citenamefont {Zlosnik}, \citenamefont {Bourliot}, \citenamefont {Ferreira},\
  and\ \citenamefont {Starkman}}]{Zuntz2010.PRD.81.104015}%
  \BibitemOpen
  \bibfield  {author} {\bibinfo {author} {\bibfnamefont {J.}~\bibnamefont
  {Zuntz}}, \bibinfo {author} {\bibfnamefont {T.~G.}\ \bibnamefont {Zlosnik}},
  \bibinfo {author} {\bibfnamefont {F.}~\bibnamefont {Bourliot}}, \bibinfo
  {author} {\bibfnamefont {P.~G.}\ \bibnamefont {Ferreira}},\ and\ \bibinfo
  {author} {\bibfnamefont {G.~D.}\ \bibnamefont {Starkman}},\ }\bibinfo {title}
  {{Vector field models of modified gravity and the dark sector}},\ \href
  {https://doi.org/10.1103/PhysRevD.81.104015} {\bibfield  {journal} {\bibinfo
  {journal} {\PRD}\ }\textbf {\bibinfo {volume} {81}},\ \bibinfo {pages}
  {104015} (\bibinfo {year} {2010})}\BibitemShut {NoStop}%
\bibitem [{\citenamefont {Xu}\ \emph {et~al.}(2015)\citenamefont {Xu},
  \citenamefont {Wang},\ and\ \citenamefont {Zhang}}]{Xu2015.PRD.92.083505}%
  \BibitemOpen
  \bibfield  {author} {\bibinfo {author} {\bibfnamefont {X.-d.}\ \bibnamefont
  {Xu}}, \bibinfo {author} {\bibfnamefont {B.}~\bibnamefont {Wang}},\ and\
  \bibinfo {author} {\bibfnamefont {P.}~\bibnamefont {Zhang}},\ }\bibinfo
  {title} {{Testing the tensor-vector-scalar theory with the latest
  cosmological observations}},\ \href
  {https://doi.org/10.1103/PhysRevD.92.083505} {\bibfield  {journal} {\bibinfo
  {journal} {\PRD}\ }\textbf {\bibinfo {volume} {92}},\ \bibinfo {pages}
  {083505} (\bibinfo {year} {2015})}\BibitemShut {NoStop}%
\bibitem [{\citenamefont {{Tan}}\ and\ \citenamefont
  {{Woodard}}(2018)}]{Tan2018.JCAP.05.037}%
  \BibitemOpen
  \bibfield  {author} {\bibinfo {author} {\bibfnamefont {L.}~\bibnamefont
  {{Tan}}}\ and\ \bibinfo {author} {\bibfnamefont {R.~P.}\ \bibnamefont
  {{Woodard}}},\ }\bibinfo {title} {{Structure formation in nonlocal MOND}},\
  \href {https://doi.org/10.1088/1475-7516/2018/05/037} {\bibfield  {journal}
  {\bibinfo  {journal} {\JCAP}\ }\bibfield  {number} {\bibinfo  {number} {05}}
  (2018)\ 037}\BibitemShut {NoStop}%
\bibitem [{\citenamefont {Skordis}\ and\ \citenamefont
  {Z\l{}o\'{s}nik}(2021)}]{Skordis2021.PRL.127.161302}%
  \BibitemOpen
  \bibfield  {author} {\bibinfo {author} {\bibfnamefont {C.}~\bibnamefont
  {Skordis}}\ and\ \bibinfo {author} {\bibfnamefont {T.}~\bibnamefont
  {Z\l{}o\'{s}nik}},\ }\bibinfo {title} {{New Relativistic Theory for Modified
  Newtonian Dynamics}},\ \href {https://doi.org/10.1103/PhysRevLett.127.161302}
  {\bibfield  {journal} {\bibinfo  {journal} {\PRL}\ }\textbf {\bibinfo
  {volume} {127}},\ \bibinfo {pages} {161302} (\bibinfo {year}
  {2021})}\BibitemShut {NoStop}%
\bibitem [{\citenamefont {{Milgrom}}(1994)}]{Milgrom1994.AnnPhysNY.229.384}%
  \BibitemOpen
  \bibfield  {author} {\bibinfo {author} {\bibfnamefont {M.}~\bibnamefont
  {{Milgrom}}},\ }\bibinfo {title} {{Dynamics with a nonstandard
  inertia-acceleration relation: An alternative to dark matter in galactic
  systems}},\ \href {https://doi.org/10.1006/aphy.1994.1012} {\bibfield
  {journal} {\bibinfo  {journal} {\AnnPhysNY}\ }\textbf {\bibinfo {volume}
  {229}},\ \bibinfo {pages} {384} (\bibinfo {year} {1994})}\BibitemShut
  {NoStop}%
\bibitem [{\citenamefont {{Milgrom}}(1999)}]{Milgrom1999.PhysLettA.253.273}%
  \BibitemOpen
  \bibfield  {author} {\bibinfo {author} {\bibfnamefont {M.}~\bibnamefont
  {{Milgrom}}},\ }\bibinfo {title} {{The modified dynamics as a vacuum
  effect}},\ \href {https://doi.org/10.1016/S0375-9601(99)00077-8} {\bibfield
  {journal} {\bibinfo  {journal} {\PhysLettA}\ }\textbf {\bibinfo {volume}
  {253}},\ \bibinfo {pages} {273} (\bibinfo {year} {1999})}\BibitemShut
  {NoStop}%
\bibitem [{\citenamefont {{Petersen}}\ and\ \citenamefont
  {{Lelli}}(2020)}]{Petersen2020.AaA.636.A56}%
  \BibitemOpen
  \bibfield  {author} {\bibinfo {author} {\bibfnamefont {J.}~\bibnamefont
  {{Petersen}}}\ and\ \bibinfo {author} {\bibfnamefont {F.}~\bibnamefont
  {{Lelli}}},\ }\bibinfo {title} {{A first attempt to differentiate between
  modified gravity and modified inertia with galaxy rotation curves}},\ \href
  {https://doi.org/10.1051/0004-6361/201936964} {\bibfield  {journal} {\bibinfo
   {journal} {\AaA}\ }\textbf {\bibinfo {volume} {636}},\ \bibinfo {eid} {A56}
  (\bibinfo {year} {2020})}\BibitemShut {NoStop}%
\bibitem [{\citenamefont {Milgrom}(2022)}]{Milgrom2022.PRD.106.064060}%
  \BibitemOpen
  \bibfield  {author} {\bibinfo {author} {\bibfnamefont {M.}~\bibnamefont
  {Milgrom}},\ }\bibinfo {title} {{Models of a modified-inertia formulation of
  MOND}},\ \href {https://doi.org/10.1103/PhysRevD.106.064060} {\bibfield
  {journal} {\bibinfo  {journal} {\PRD}\ }\textbf {\bibinfo {volume} {106}},\
  \bibinfo {pages} {064060} (\bibinfo {year} {2022})}\BibitemShut {NoStop}%
\bibitem [{\citenamefont {McGaugh}\ \emph {et~al.}(2016)\citenamefont
  {McGaugh}, \citenamefont {Lelli},\ and\ \citenamefont
  {Schombert}}]{McGaugh2016.PRL.117.201101}%
  \BibitemOpen
  \bibfield  {author} {\bibinfo {author} {\bibfnamefont {S.~S.}\ \bibnamefont
  {McGaugh}}, \bibinfo {author} {\bibfnamefont {F.}~\bibnamefont {Lelli}},\
  and\ \bibinfo {author} {\bibfnamefont {J.~M.}\ \bibnamefont {Schombert}},\
  }\bibinfo {title} {{Radial Acceleration Relation in Rotationally Supported
  Galaxies}},\ \href {https://doi.org/10.1103/PhysRevLett.117.201101}
  {\bibfield  {journal} {\bibinfo  {journal} {\PRL}\ }\textbf {\bibinfo
  {volume} {117}},\ \bibinfo {pages} {201101} (\bibinfo {year}
  {2016})}\BibitemShut {NoStop}%
\bibitem [{\citenamefont {{Lelli}}\ \emph {et~al.}(2017)\citenamefont
  {{Lelli}}, \citenamefont {{McGaugh}}, \citenamefont {{Schombert}},\ and\
  \citenamefont {{Pawlowski}}}]{Lelli2017.ApJ.836.152}%
  \BibitemOpen
  \bibfield  {author} {\bibinfo {author} {\bibfnamefont {F.}~\bibnamefont
  {{Lelli}}}, \bibinfo {author} {\bibfnamefont {S.~S.}\ \bibnamefont
  {{McGaugh}}}, \bibinfo {author} {\bibfnamefont {J.~M.}\ \bibnamefont
  {{Schombert}}},\ and\ \bibinfo {author} {\bibfnamefont {M.~S.}\ \bibnamefont
  {{Pawlowski}}},\ }\bibinfo {title} {{One law to rule them all: The radial
  acceleration relation of galaxies}},\ \href
  {https://doi.org/10.3847/1538-4357/836/2/152} {\bibfield  {journal} {\bibinfo
   {journal} {\ApJ}\ }\textbf {\bibinfo {volume} {836}},\ \bibinfo {eid} {152}
  (\bibinfo {year} {2017})}\BibitemShut {NoStop}%
\bibitem [{\citenamefont {{Tian}}\ \emph {et~al.}(2020)\citenamefont {{Tian}},
  \citenamefont {{Umetsu}}, \citenamefont {{Ko}}, \citenamefont {{Donahue}},\
  and\ \citenamefont {{Chiu}}}]{Tian2020.ApJ.896.70}%
  \BibitemOpen
  \bibfield  {author} {\bibinfo {author} {\bibfnamefont {Y.}~\bibnamefont
  {{Tian}}}, \bibinfo {author} {\bibfnamefont {K.}~\bibnamefont {{Umetsu}}},
  \bibinfo {author} {\bibfnamefont {C.-M.}\ \bibnamefont {{Ko}}}, \bibinfo
  {author} {\bibfnamefont {M.}~\bibnamefont {{Donahue}}},\ and\ \bibinfo
  {author} {\bibfnamefont {I.~N.}\ \bibnamefont {{Chiu}}},\ }\bibinfo {title}
  {{The radial acceleration relation in CLASH galaxy clusters}},\ \href
  {https://doi.org/10.3847/1538-4357/ab8e3d} {\bibfield  {journal} {\bibinfo
  {journal} {\ApJ}\ }\textbf {\bibinfo {volume} {896}},\ \bibinfo {eid} {70}
  (\bibinfo {year} {2020})}\BibitemShut {NoStop}%
\bibitem [{\citenamefont {Dai}\ and\ \citenamefont
  {Lu}(2017)}]{Dai2017.PRD.96.124016}%
  \BibitemOpen
  \bibfield  {author} {\bibinfo {author} {\bibfnamefont {D.-C.}\ \bibnamefont
  {Dai}}\ and\ \bibinfo {author} {\bibfnamefont {C.}~\bibnamefont {Lu}},\
  }\bibinfo {title} {{Can the $\Lambda$CDM model reproduce MOND-like
  behavior?}},\ \href {https://doi.org/10.1103/PhysRevD.96.124016} {\bibfield
  {journal} {\bibinfo  {journal} {\PRD}\ }\textbf {\bibinfo {volume} {96}},\
  \bibinfo {pages} {124016} (\bibinfo {year} {2017})}\BibitemShut {NoStop}%
\bibitem [{\citenamefont {{Keller}}\ and\ \citenamefont
  {{Wadsley}}(2017)}]{Keller2017.ApJL.835.L17}%
  \BibitemOpen
  \bibfield  {author} {\bibinfo {author} {\bibfnamefont {B.~W.}\ \bibnamefont
  {{Keller}}}\ and\ \bibinfo {author} {\bibfnamefont {J.~W.}\ \bibnamefont
  {{Wadsley}}},\ }\bibinfo {title} {{$\Lambda$CDM is consistent with SPARC
  radial acceleration relation}},\ \href
  {https://doi.org/10.3847/2041-8213/835/1/L17} {\bibfield  {journal} {\bibinfo
   {journal} {\ApJL}\ }\textbf {\bibinfo {volume} {835}},\ \bibinfo {eid} {L17}
  (\bibinfo {year} {2017})}\BibitemShut {NoStop}%
\bibitem [{\citenamefont {Garaldi}\ \emph {et~al.}(2018)\citenamefont
  {Garaldi}, \citenamefont {Romano-D\'{\i}az}, \citenamefont {Porciani},\ and\
  \citenamefont {Pawlowski}}]{Garaldi2018.PRL.120.261301}%
  \BibitemOpen
  \bibfield  {author} {\bibinfo {author} {\bibfnamefont {E.}~\bibnamefont
  {Garaldi}}, \bibinfo {author} {\bibfnamefont {E.}~\bibnamefont
  {Romano-D\'{\i}az}}, \bibinfo {author} {\bibfnamefont {C.}~\bibnamefont
  {Porciani}},\ and\ \bibinfo {author} {\bibfnamefont {M.~S.}\ \bibnamefont
  {Pawlowski}},\ }\bibinfo {title} {{Radial Acceleration Relation of
  $\Lambda$CDM Satellite Galaxies}},\ \href
  {https://doi.org/10.1103/PhysRevLett.120.261301} {\bibfield  {journal}
  {\bibinfo  {journal} {\PRL}\ }\textbf {\bibinfo {volume} {120}},\ \bibinfo
  {pages} {261301} (\bibinfo {year} {2018})}\BibitemShut {NoStop}%
\bibitem [{\citenamefont {{Dutton}}\ \emph {et~al.}(2019)\citenamefont
  {{Dutton}}, \citenamefont {{Macci{\`o}}}, \citenamefont {{Obreja}},\ and\
  \citenamefont {{Buck}}}]{Dutton2019.MNRAS.485.1886}%
  \BibitemOpen
  \bibfield  {author} {\bibinfo {author} {\bibfnamefont {A.~A.}\ \bibnamefont
  {{Dutton}}}, \bibinfo {author} {\bibfnamefont {A.~V.}\ \bibnamefont
  {{Macci{\`o}}}}, \bibinfo {author} {\bibfnamefont {A.}~\bibnamefont
  {{Obreja}}},\ and\ \bibinfo {author} {\bibfnamefont {T.}~\bibnamefont
  {{Buck}}},\ }\bibinfo {title} {{NIHAO -- XVIII. Origin of the MOND
  phenomenology of galactic rotation curves in a $\Lambda$CDM universe}},\
  \href {https://doi.org/10.1093/mnras/stz531} {\bibfield  {journal} {\bibinfo
  {journal} {\MNRAS}\ }\textbf {\bibinfo {volume} {485}},\ \bibinfo {pages}
  {1886} (\bibinfo {year} {2019})}\BibitemShut {NoStop}%
\bibitem [{\citenamefont {{Durazo}}\ \emph {et~al.}(2017)\citenamefont
  {{Durazo}}, \citenamefont {{Hernandez}}, \citenamefont {{Cervantes Sodi}},\
  and\ \citenamefont {{S{\'a}nchez}}}]{Durazo2017.ApJ.837.179}%
  \BibitemOpen
  \bibfield  {author} {\bibinfo {author} {\bibfnamefont {R.}~\bibnamefont
  {{Durazo}}}, \bibinfo {author} {\bibfnamefont {X.}~\bibnamefont
  {{Hernandez}}}, \bibinfo {author} {\bibfnamefont {B.}~\bibnamefont
  {{Cervantes Sodi}}},\ and\ \bibinfo {author} {\bibfnamefont {S.~F.}\
  \bibnamefont {{S{\'a}nchez}}},\ }\bibinfo {title} {{A universal velocity
  dispersion profile for pressure supported systems: Evidence for MONDian
  gravity across seven orders of magnitude in mass}},\ \href
  {https://doi.org/10.3847/1538-4357/aa619a} {\bibfield  {journal} {\bibinfo
  {journal} {\ApJ}\ }\textbf {\bibinfo {volume} {837}},\ \bibinfo {eid} {179}
  (\bibinfo {year} {2017})}\BibitemShut {NoStop}%
\bibitem [{\citenamefont {{Navarro}}\ \emph {et~al.}(2017)\citenamefont
  {{Navarro}}, \citenamefont {{Ben{\'\i}tez-Llambay}}, \citenamefont
  {{Fattahi}}, \citenamefont {{Frenk}}, \citenamefont {{Ludlow}}, \citenamefont
  {{Oman}}, \citenamefont {{Schaller}},\ and\ \citenamefont
  {{Theuns}}}]{Navarro2017.MNRAS.471.1841}%
  \BibitemOpen
  \bibfield  {author} {\bibinfo {author} {\bibfnamefont {J.~F.}\ \bibnamefont
  {{Navarro}}}, \bibinfo {author} {\bibfnamefont {A.}~\bibnamefont
  {{Ben{\'\i}tez-Llambay}}}, \bibinfo {author} {\bibfnamefont {A.}~\bibnamefont
  {{Fattahi}}}, \bibinfo {author} {\bibfnamefont {C.~S.}\ \bibnamefont
  {{Frenk}}}, \bibinfo {author} {\bibfnamefont {A.~D.}\ \bibnamefont
  {{Ludlow}}}, \bibinfo {author} {\bibfnamefont {K.~A.}\ \bibnamefont
  {{Oman}}}, \bibinfo {author} {\bibfnamefont {M.}~\bibnamefont {{Schaller}}},\
  and\ \bibinfo {author} {\bibfnamefont {T.}~\bibnamefont {{Theuns}}},\
  }\bibinfo {title} {{The origin of the mass discrepancy-acceleration relation
  in $\Lambda$CDM}},\ \href {https://doi.org/10.1093/mnras/stx1705} {\bibfield
  {journal} {\bibinfo  {journal} {\MNRAS}\ }\textbf {\bibinfo {volume} {471}},\
  \bibinfo {pages} {1841} (\bibinfo {year} {2017})}\BibitemShut {NoStop}%
\bibitem [{\citenamefont {{Ludlow}}\ \emph {et~al.}(2017)\citenamefont
  {{Ludlow}} \emph {et~al.}}]{Ludlow2017.PRL.118.161103}%
  \BibitemOpen
  \bibfield  {author} {\bibinfo {author} {\bibfnamefont {A.~D.}\ \bibnamefont
  {{Ludlow}}} \emph {et~al.},\ }\bibinfo {title} {{Mass-Discrepancy
  Acceleration Relation: A Natural Outcome of Galaxy Formation in Cold Dark
  Matter Halos}},\ \href {https://doi.org/10.1103/PhysRevLett.118.161103}
  {\bibfield  {journal} {\bibinfo  {journal} {\PRL}\ }\textbf {\bibinfo
  {volume} {118}},\ \bibinfo {eid} {161103} (\bibinfo {year}
  {2017})}\BibitemShut {NoStop}%
\bibitem [{\citenamefont {{Mayer}}\ \emph {et~al.}(2023)\citenamefont
  {{Mayer}}, \citenamefont {{Teklu}}, \citenamefont {{Dolag}},\ and\
  \citenamefont {{Remus}}}]{Mayer2023.MNRAS.518.257}%
  \BibitemOpen
  \bibfield  {author} {\bibinfo {author} {\bibfnamefont {A.~C.}\ \bibnamefont
  {{Mayer}}}, \bibinfo {author} {\bibfnamefont {A.~F.}\ \bibnamefont
  {{Teklu}}}, \bibinfo {author} {\bibfnamefont {K.}~\bibnamefont {{Dolag}}},\
  and\ \bibinfo {author} {\bibfnamefont {R.-S.}\ \bibnamefont {{Remus}}},\
  }\bibinfo {title} {{$\Lambda$CDM with baryons versus MOND: The time evolution
  of the universal acceleration scale in the \textit{Magneticum}
  simulations}},\ \href {https://doi.org/10.1093/mnras/stac3017} {\bibfield
  {journal} {\bibinfo  {journal} {\MNRAS}\ }\textbf {\bibinfo {volume} {518}},\
  \bibinfo {pages} {257} (\bibinfo {year} {2023})}\BibitemShut {NoStop}%
\bibitem [{\citenamefont {Bekenstein}\ and\ \citenamefont
  {Sagi}(2008)}]{Bekenstein2008.PRD.77.103512}%
  \BibitemOpen
  \bibfield  {author} {\bibinfo {author} {\bibfnamefont {J.~D.}\ \bibnamefont
  {Bekenstein}}\ and\ \bibinfo {author} {\bibfnamefont {E.}~\bibnamefont
  {Sagi}},\ }\bibinfo {title} {{Do Newton's $G$ and Milgrom's ${a}_{0}$ vary
  with cosmological epoch?}},\ \href
  {https://doi.org/10.1103/PhysRevD.77.103512} {\bibfield  {journal} {\bibinfo
  {journal} {\PRD}\ }\textbf {\bibinfo {volume} {77}},\ \bibinfo {pages}
  {103512} (\bibinfo {year} {2008})}\BibitemShut {NoStop}%
\bibitem [{\citenamefont {Abbott}\ \emph {et~al.}(2016)\citenamefont {Abbott}
  \emph {et~al.}}]{Abbott2016.PRL.116.061102}%
  \BibitemOpen
  \bibfield  {author} {\bibinfo {author} {\bibfnamefont {B.~P.}\ \bibnamefont
  {Abbott}} \emph {et~al.} (\bibinfo {collaboration} {LIGO Scientific
  Collaboration and Virgo Collaboration}),\ }\bibinfo {title} {{Observation of
  Gravitational Waves from a Binary Black Hole Merger}},\ \href
  {https://doi.org/10.1103/PhysRevLett.116.061102} {\bibfield  {journal}
  {\bibinfo  {journal} {\PRL}\ }\textbf {\bibinfo {volume} {116}},\ \bibinfo
  {pages} {061102} (\bibinfo {year} {2016})}\BibitemShut {NoStop}%
\bibitem [{\citenamefont {Abbott}\ \emph
  {et~al.}(2017{\natexlab{a}})\citenamefont {Abbott} \emph
  {et~al.}}]{Abbott2017.PRL.119.141101}%
  \BibitemOpen
  \bibfield  {author} {\bibinfo {author} {\bibfnamefont {B.~P.}\ \bibnamefont
  {Abbott}} \emph {et~al.} (\bibinfo {collaboration} {LIGO Scientific
  Collaboration and Virgo Collaboration}),\ }\bibinfo {title} {{GW170814: A
  Three-Detector Observation of Gravitational Waves from a Binary Black Hole
  Coalescence}},\ \href {https://doi.org/10.1103/PhysRevLett.119.141101}
  {\bibfield  {journal} {\bibinfo  {journal} {\PRL}\ }\textbf {\bibinfo
  {volume} {119}},\ \bibinfo {pages} {141101} (\bibinfo {year}
  {2017}{\natexlab{a}})}\BibitemShut {NoStop}%
\bibitem [{\citenamefont {Takeda}\ \emph {et~al.}(2021)\citenamefont {Takeda},
  \citenamefont {Morisaki},\ and\ \citenamefont
  {Nishizawa}}]{Takeda2021.PRD.103.064037}%
  \BibitemOpen
  \bibfield  {author} {\bibinfo {author} {\bibfnamefont {H.}~\bibnamefont
  {Takeda}}, \bibinfo {author} {\bibfnamefont {S.}~\bibnamefont {Morisaki}},\
  and\ \bibinfo {author} {\bibfnamefont {A.}~\bibnamefont {Nishizawa}},\
  }\bibinfo {title} {{Pure polarization test of GW170814 and GW170817 using
  waveforms consistent with modified theories of gravity}},\ \href
  {https://doi.org/10.1103/PhysRevD.103.064037} {\bibfield  {journal} {\bibinfo
   {journal} {\PRD}\ }\textbf {\bibinfo {volume} {103}},\ \bibinfo {pages}
  {064037} (\bibinfo {year} {2021})}\BibitemShut {NoStop}%
\bibitem [{\citenamefont {Abbott}\ \emph
  {et~al.}(2017{\natexlab{b}})\citenamefont {Abbott} \emph
  {et~al.}}]{Abbott2017.PRL.119.161101}%
  \BibitemOpen
  \bibfield  {author} {\bibinfo {author} {\bibfnamefont {B.~P.}\ \bibnamefont
  {Abbott}} \emph {et~al.} (\bibinfo {collaboration} {LIGO Scientific
  Collaboration and Virgo Collaboration}),\ }\bibinfo {title} {{GW170817:
  Observation of Gravitational Waves from a Binary Neutron Star Inspiral}},\
  \href {https://doi.org/10.1103/PhysRevLett.119.161101} {\bibfield  {journal}
  {\bibinfo  {journal} {\PRL}\ }\textbf {\bibinfo {volume} {119}},\ \bibinfo
  {pages} {161101} (\bibinfo {year} {2017}{\natexlab{b}})}\BibitemShut
  {NoStop}%
\bibitem [{\citenamefont {Aghanim}\ \emph {et~al.}(2020)\citenamefont {Aghanim}
  \emph {et~al.}}]{Aghanim2020.AaA.641.A6}%
  \BibitemOpen
  \bibfield  {author} {\bibinfo {author} {\bibfnamefont {N.}~\bibnamefont
  {Aghanim}} \emph {et~al.} (\bibinfo {collaboration} {Planck Collaboration}),\
  }\bibinfo {title} {{\textit{Planck} 2018 results VI. Cosmological
  parameters}},\ \href {https://doi.org/10.1051/0004-6361/201833910} {\bibfield
   {journal} {\bibinfo  {journal} {\AaA}\ }\textbf {\bibinfo {volume} {641}},\
  \bibinfo {pages} {A6} (\bibinfo {year} {2020})}\BibitemShut {NoStop}%
\bibitem [{\citenamefont {Brans}\ and\ \citenamefont
  {Dicke}(1961)}]{Brans1961.PhysRev.124.925}%
  \BibitemOpen
  \bibfield  {author} {\bibinfo {author} {\bibfnamefont {C.}~\bibnamefont
  {Brans}}\ and\ \bibinfo {author} {\bibfnamefont {R.~H.}\ \bibnamefont
  {Dicke}},\ }\bibinfo {title} {{Mach's principle and a relativistic theory of
  gravitation}},\ \href {https://doi.org/10.1103/PhysRev.124.925} {\bibfield
  {journal} {\bibinfo  {journal} {\PhysRev}\ }\textbf {\bibinfo {volume}
  {124}},\ \bibinfo {pages} {925} (\bibinfo {year} {1961})}\BibitemShut
  {NoStop}%
\bibitem [{\citenamefont {Damour}\ \emph {et~al.}(1990)\citenamefont {Damour},
  \citenamefont {Gibbons},\ and\ \citenamefont
  {Gundlach}}]{Damour1990.PRL.64.123}%
  \BibitemOpen
  \bibfield  {author} {\bibinfo {author} {\bibfnamefont {T.}~\bibnamefont
  {Damour}}, \bibinfo {author} {\bibfnamefont {G.~W.}\ \bibnamefont
  {Gibbons}},\ and\ \bibinfo {author} {\bibfnamefont {C.}~\bibnamefont
  {Gundlach}},\ }\bibinfo {title} {{Dark Matter, Time-Varying $G$, and a
  Dilaton Field}},\ \href {https://doi.org/10.1103/PhysRevLett.64.123}
  {\bibfield  {journal} {\bibinfo  {journal} {\PRL}\ }\textbf {\bibinfo
  {volume} {64}},\ \bibinfo {pages} {123} (\bibinfo {year} {1990})}\BibitemShut
  {NoStop}%
\bibitem [{\citenamefont {Babichev}\ \emph {et~al.}(2011)\citenamefont
  {Babichev}, \citenamefont {Deffayet},\ and\ \citenamefont
  {Esposito-Far\`ese}}]{Babichev2011.PRL.107.251102}%
  \BibitemOpen
  \bibfield  {author} {\bibinfo {author} {\bibfnamefont {E.}~\bibnamefont
  {Babichev}}, \bibinfo {author} {\bibfnamefont {C.}~\bibnamefont {Deffayet}},\
  and\ \bibinfo {author} {\bibfnamefont {G.}~\bibnamefont
  {Esposito-Far\`ese}},\ }\bibinfo {title} {{Constraints on Shift-Symmetric
  Scalar-Tensor Theories with a Vainshtein Mechanism from Bounds on the Time
  Variation of $G$}},\ \href {https://doi.org/10.1103/PhysRevLett.107.251102}
  {\bibfield  {journal} {\bibinfo  {journal} {\PRL}\ }\textbf {\bibinfo
  {volume} {107}},\ \bibinfo {pages} {251102} (\bibinfo {year}
  {2011})}\BibitemShut {NoStop}%
\bibitem [{\citenamefont {Zhang}\ \emph {et~al.}(2019)\citenamefont {Zhang},
  \citenamefont {Niu},\ and\ \citenamefont {Zhao}}]{Zhang2019.PRD.100.024038}%
  \BibitemOpen
  \bibfield  {author} {\bibinfo {author} {\bibfnamefont {X.}~\bibnamefont
  {Zhang}}, \bibinfo {author} {\bibfnamefont {R.}~\bibnamefont {Niu}},\ and\
  \bibinfo {author} {\bibfnamefont {W.}~\bibnamefont {Zhao}},\ }\bibinfo
  {title} {{Constraining the scalar-tensor gravity theories with and without
  screening mechanisms by combined observations}},\ \href
  {https://doi.org/10.1103/PhysRevD.100.024038} {\bibfield  {journal} {\bibinfo
   {journal} {\PRD}\ }\textbf {\bibinfo {volume} {100}},\ \bibinfo {pages}
  {024038} (\bibinfo {year} {2019})}\BibitemShut {NoStop}%
\bibitem [{\citenamefont {{Burrage}}\ and\ \citenamefont
  {{Dombrowski}}(2020)}]{Burrage2020.JCAP.07.060}%
  \BibitemOpen
  \bibfield  {author} {\bibinfo {author} {\bibfnamefont {C.}~\bibnamefont
  {{Burrage}}}\ and\ \bibinfo {author} {\bibfnamefont {J.}~\bibnamefont
  {{Dombrowski}}},\ }\bibinfo {title} {{Constraining the cosmological evolution
  of scalar-tensor theories with local measurements of the time variation of
  $G$}},\ \href {https://doi.org/10.1088/1475-7516/2020/07/060} {\bibfield
  {journal} {\bibinfo  {journal} {\JCAP}\ }\bibfield  {number} {\bibinfo
  {number} {07}} (2020)\ 060}\BibitemShut {NoStop}%
\bibitem [{\citenamefont {{Barreira}}\ \emph {et~al.}(2014)\citenamefont
  {{Barreira}}, \citenamefont {{Li}}, \citenamefont {{Hellwing}}, \citenamefont
  {{Baugh}},\ and\ \citenamefont {{Pascoli}}}]{Barreira2014.JCAP.09.031}%
  \BibitemOpen
  \bibfield  {author} {\bibinfo {author} {\bibfnamefont {A.}~\bibnamefont
  {{Barreira}}}, \bibinfo {author} {\bibfnamefont {B.}~\bibnamefont {{Li}}},
  \bibinfo {author} {\bibfnamefont {W.~A.}\ \bibnamefont {{Hellwing}}},
  \bibinfo {author} {\bibfnamefont {C.~M.}\ \bibnamefont {{Baugh}}},\ and\
  \bibinfo {author} {\bibfnamefont {S.}~\bibnamefont {{Pascoli}}},\ }\bibinfo
  {title} {{Nonlinear structure formation in nonlocal gravity}},\ \href
  {https://doi.org/10.1088/1475-7516/2014/09/031} {\bibfield  {journal}
  {\bibinfo  {journal} {\JCAP}\ }\bibfield  {number} {\bibinfo  {number} {09}}
  (2014)\ 031}\BibitemShut {NoStop}%
\bibitem [{\citenamefont {{Belgacem}}\ \emph {et~al.}(2019)\citenamefont
  {{Belgacem}}, \citenamefont {{Finke}}, \citenamefont {{Frassino}},\ and\
  \citenamefont {{Maggiore}}}]{Belgacem2019.JCAP.02.035}%
  \BibitemOpen
  \bibfield  {author} {\bibinfo {author} {\bibfnamefont {E.}~\bibnamefont
  {{Belgacem}}}, \bibinfo {author} {\bibfnamefont {A.}~\bibnamefont {{Finke}}},
  \bibinfo {author} {\bibfnamefont {A.}~\bibnamefont {{Frassino}}},\ and\
  \bibinfo {author} {\bibfnamefont {M.}~\bibnamefont {{Maggiore}}},\ }\bibinfo
  {title} {{Testing nonlocal gravity with lunar laser ranging}},\ \href
  {https://doi.org/10.1088/1475-7516/2019/02/035} {\bibfield  {journal}
  {\bibinfo  {journal} {\JCAP}\ }\bibfield  {number} {\bibinfo  {number} {02}}
  (2019)\ 035}\BibitemShut {NoStop}%
\bibitem [{\citenamefont {Tian}\ and\ \citenamefont
  {Zhu}(2019)}]{Tian2019.PRD.99.064044}%
  \BibitemOpen
  \bibfield  {author} {\bibinfo {author} {\bibfnamefont {S.~X.}\ \bibnamefont
  {Tian}}\ and\ \bibinfo {author} {\bibfnamefont {Z.-H.}\ \bibnamefont {Zhu}},\
  }\bibinfo {title} {{Newtonian approximation and possible time-varying $G$ in
  nonlocal gravities}},\ \href {https://doi.org/10.1103/PhysRevD.99.064044}
  {\bibfield  {journal} {\bibinfo  {journal} {\PRD}\ }\textbf {\bibinfo
  {volume} {99}},\ \bibinfo {pages} {064044} (\bibinfo {year}
  {2019})}\BibitemShut {NoStop}%
\bibitem [{\citenamefont {Dodelson}\ and\ \citenamefont
  {Schmidt}(2020)}]{Dodelson2020.book}%
  \BibitemOpen
  \bibfield  {author} {\bibinfo {author} {\bibfnamefont {S.}~\bibnamefont
  {Dodelson}}\ and\ \bibinfo {author} {\bibfnamefont {F.}~\bibnamefont
  {Schmidt}},\ }\href@noop {} {\emph {\bibinfo {title} {{Modern Cosmology}}}},\
  \bibinfo {edition} {2nd}\ ed.\ (\bibinfo  {publisher} {Academic Press},\
  \bibinfo {address} {London},\ \bibinfo {year} {2020})\BibitemShut {NoStop}%
\bibitem [{Note1()}]{Note1}%
  \BibitemOpen
  \bibinfo {note} {Here $\protect \mathcal {Q}_c$ is exactly the $\protect
  \mathcal {Q}_0$ used in \cite {Skordis2021.PRL.127.161302}. We do this
  replacement because the subscript $0$ indicates $z=0$ in our
  conventions.}\BibitemShut {Stop}%
\bibitem [{Note2()}]{Note2}%
  \BibitemOpen
  \bibinfo {note} {Here \protect \textit {strong} means the acceleration is
  much larger than $a_\protect \textsc {mond}$. The latter \protect \textit
  {weak} indicates the opposite case.}\BibitemShut {Stop}%
\bibitem [{\citenamefont {{Ostriker}}(1993)}]{Ostriker1993.ARAA.31.689}%
  \BibitemOpen
  \bibfield  {author} {\bibinfo {author} {\bibfnamefont {J.~P.}\ \bibnamefont
  {{Ostriker}}},\ }\bibinfo {title} {{Astronomical tests of the cold dark
  matter scenario}},\ \href
  {https://doi.org/10.1146/annurev.aa.31.090193.003353} {\bibfield  {journal}
  {\bibinfo  {journal} {\ARAA}\ }\textbf {\bibinfo {volume} {31}},\ \bibinfo
  {pages} {689} (\bibinfo {year} {1993})}\BibitemShut {NoStop}%
\bibitem [{\citenamefont {Frieman}\ \emph {et~al.}(2008)\citenamefont
  {Frieman}, \citenamefont {Turner},\ and\ \citenamefont
  {Huterer}}]{Frieman2008.ARAA.46.385}%
  \BibitemOpen
  \bibfield  {author} {\bibinfo {author} {\bibfnamefont {J.~A.}\ \bibnamefont
  {Frieman}}, \bibinfo {author} {\bibfnamefont {M.~S.}\ \bibnamefont
  {Turner}},\ and\ \bibinfo {author} {\bibfnamefont {D.}~\bibnamefont
  {Huterer}},\ }\bibinfo {title} {{Dark energy and the accelerating
  universe}},\ \href {https://doi.org/10.1146/annurev.astro.46.060407.145243}
  {\bibfield  {journal} {\bibinfo  {journal} {\ARAA}\ }\textbf {\bibinfo
  {volume} {46}},\ \bibinfo {pages} {385} (\bibinfo {year} {2008})}\BibitemShut
  {NoStop}%
\bibitem [{\citenamefont {Bertone}\ and\ \citenamefont
  {Hooper}(2018)}]{Bertone2018.RMP.90.045002}%
  \BibitemOpen
  \bibfield  {author} {\bibinfo {author} {\bibfnamefont {G.}~\bibnamefont
  {Bertone}}\ and\ \bibinfo {author} {\bibfnamefont {D.}~\bibnamefont
  {Hooper}},\ }\bibinfo {title} {{History of dark matter}},\ \href
  {https://doi.org/10.1103/RevModPhys.90.045002} {\bibfield  {journal}
  {\bibinfo  {journal} {\RMP}\ }\textbf {\bibinfo {volume} {90}},\ \bibinfo
  {pages} {045002} (\bibinfo {year} {2018})}\BibitemShut {NoStop}%
\bibitem [{\citenamefont {{Penzo}}\ \emph {et~al.}(2014)\citenamefont
  {{Penzo}}, \citenamefont {{Macci{\`o}}}, \citenamefont {{Casarini}},
  \citenamefont {{Stinson}},\ and\ \citenamefont
  {{Wadsley}}}]{Penzo2014.MNRAS.442.176}%
  \BibitemOpen
  \bibfield  {author} {\bibinfo {author} {\bibfnamefont {C.}~\bibnamefont
  {{Penzo}}}, \bibinfo {author} {\bibfnamefont {A.~V.}\ \bibnamefont
  {{Macci{\`o}}}}, \bibinfo {author} {\bibfnamefont {L.}~\bibnamefont
  {{Casarini}}}, \bibinfo {author} {\bibfnamefont {G.~S.}\ \bibnamefont
  {{Stinson}}},\ and\ \bibinfo {author} {\bibfnamefont {J.}~\bibnamefont
  {{Wadsley}}},\ }\bibinfo {title} {{Dark MaGICC: The effect of dark energy on
  disc galaxy formation. Cosmology does matter}},\ \href
  {https://doi.org/10.1093/mnras/stu857} {\bibfield  {journal} {\bibinfo
  {journal} {\MNRAS}\ }\textbf {\bibinfo {volume} {442}},\ \bibinfo {pages}
  {176} (\bibinfo {year} {2014})}\BibitemShut {NoStop}%
\bibitem [{\citenamefont {Schive}\ \emph {et~al.}(2014)\citenamefont {Schive},
  \citenamefont {Chiueh},\ and\ \citenamefont
  {Broadhurst}}]{Schive2014.NatPhys.10.496}%
  \BibitemOpen
  \bibfield  {author} {\bibinfo {author} {\bibfnamefont {H.-Y.}\ \bibnamefont
  {Schive}}, \bibinfo {author} {\bibfnamefont {T.}~\bibnamefont {Chiueh}},\
  and\ \bibinfo {author} {\bibfnamefont {T.}~\bibnamefont {Broadhurst}},\
  }\bibinfo {title} {{Cosmic structure as the quantum interference of a
  coherent dark wave}},\ \href {https://doi.org/10.1038/nphys2996} {\bibfield
  {journal} {\bibinfo  {journal} {\NatPhys}\ }\textbf {\bibinfo {volume}
  {10}},\ \bibinfo {pages} {496} (\bibinfo {year} {2014})}\BibitemShut
  {NoStop}%
\bibitem [{\citenamefont {Williams}\ \emph {et~al.}(2004)\citenamefont
  {Williams}, \citenamefont {Turyshev},\ and\ \citenamefont
  {Boggs}}]{Williams2004.PRL.93.261101}%
  \BibitemOpen
  \bibfield  {author} {\bibinfo {author} {\bibfnamefont {J.~G.}\ \bibnamefont
  {Williams}}, \bibinfo {author} {\bibfnamefont {S.~G.}\ \bibnamefont
  {Turyshev}},\ and\ \bibinfo {author} {\bibfnamefont {D.~H.}\ \bibnamefont
  {Boggs}},\ }\bibinfo {title} {{Progress in Lunar Laser Ranging Tests of
  Relativistic Gravity}},\ \href
  {https://doi.org/10.1103/PhysRevLett.93.261101} {\bibfield  {journal}
  {\bibinfo  {journal} {\PRL}\ }\textbf {\bibinfo {volume} {93}},\ \bibinfo
  {pages} {261101} (\bibinfo {year} {2004})}\BibitemShut {NoStop}%
\bibitem [{\citenamefont {{Hofmann}}\ \emph {et~al.}(2010)\citenamefont
  {{Hofmann}}, \citenamefont {{M{\"u}ller}},\ and\ \citenamefont
  {{Biskupek}}}]{Hofmann2010.AaA.522.L5}%
  \BibitemOpen
  \bibfield  {author} {\bibinfo {author} {\bibfnamefont {F.}~\bibnamefont
  {{Hofmann}}}, \bibinfo {author} {\bibfnamefont {J.}~\bibnamefont
  {{M{\"u}ller}}},\ and\ \bibinfo {author} {\bibfnamefont {L.}~\bibnamefont
  {{Biskupek}}},\ }\bibinfo {title} {{Lunar laser ranging test of the Nordtvedt
  parameter and a possible variation in the gravitational constant}},\ \href
  {https://doi.org/10.1051/0004-6361/201015659} {\bibfield  {journal} {\bibinfo
   {journal} {\AaA}\ }\textbf {\bibinfo {volume} {522}},\ \bibinfo {eid} {L5}
  (\bibinfo {year} {2010})}\BibitemShut {NoStop}%
\bibitem [{\citenamefont {{Zhu}}\ \emph {et~al.}(2015)\citenamefont {{Zhu}}
  \emph {et~al.}}]{Zhu2015.ApJ.809.41}%
  \BibitemOpen
  \bibfield  {author} {\bibinfo {author} {\bibfnamefont {W.~W.}\ \bibnamefont
  {{Zhu}}} \emph {et~al.},\ }\bibinfo {title} {{Testing theories of gravitation
  using 21-year timing of pulsar binary J1713+0747}},\ \href
  {https://doi.org/10.1088/0004-637X/809/1/41} {\bibfield  {journal} {\bibinfo
  {journal} {\ApJ}\ }\textbf {\bibinfo {volume} {809}},\ \bibinfo {eid} {41}
  (\bibinfo {year} {2015})}\BibitemShut {NoStop}%
\bibitem [{\citenamefont {Famaey}\ \emph {et~al.}(2007)\citenamefont {Famaey},
  \citenamefont {Gentile}, \citenamefont {Bruneton},\ and\ \citenamefont
  {Zhao}}]{Famaey2007.PRD.75.063002}%
  \BibitemOpen
  \bibfield  {author} {\bibinfo {author} {\bibfnamefont {B.}~\bibnamefont
  {Famaey}}, \bibinfo {author} {\bibfnamefont {G.}~\bibnamefont {Gentile}},
  \bibinfo {author} {\bibfnamefont {J.-P.}\ \bibnamefont {Bruneton}},\ and\
  \bibinfo {author} {\bibfnamefont {H.}~\bibnamefont {Zhao}},\ }\bibinfo
  {title} {{Insight into the baryon-gravity relation in galaxies}},\ \href
  {https://doi.org/10.1103/PhysRevD.75.063002} {\bibfield  {journal} {\bibinfo
  {journal} {\PRD}\ }\textbf {\bibinfo {volume} {75}},\ \bibinfo {pages}
  {063002} (\bibinfo {year} {2007})}\BibitemShut {NoStop}%
\bibitem [{\citenamefont {Deffayet}\ \emph {et~al.}(2014)\citenamefont
  {Deffayet}, \citenamefont {Esposito-Far\`ese},\ and\ \citenamefont
  {Woodard}}]{Deffayet2014.PRD.90.064038}%
  \BibitemOpen
  \bibfield  {author} {\bibinfo {author} {\bibfnamefont {C.}~\bibnamefont
  {Deffayet}}, \bibinfo {author} {\bibfnamefont {G.}~\bibnamefont
  {Esposito-Far\`ese}},\ and\ \bibinfo {author} {\bibfnamefont {R.~P.}\
  \bibnamefont {Woodard}},\ }\bibinfo {title} {{Field equations and cosmology
  for a class of nonlocal metric models of MOND}},\ \href
  {https://doi.org/10.1103/PhysRevD.90.064038} {\bibfield  {journal} {\bibinfo
  {journal} {\PRD}\ }\textbf {\bibinfo {volume} {90}},\ \bibinfo {pages}
  {064038} (\bibinfo {year} {2014})}\BibitemShut {NoStop}%
\bibitem [{\citenamefont {Milgrom}(2015)}]{Milgrom2015.PRD.91.044009}%
  \BibitemOpen
  \bibfield  {author} {\bibinfo {author} {\bibfnamefont {M.}~\bibnamefont
  {Milgrom}},\ }\bibinfo {title} {{Cosmological variation of the MOND constant:
  Secular effects on galactic systems}},\ \href
  {https://doi.org/10.1103/PhysRevD.91.044009} {\bibfield  {journal} {\bibinfo
  {journal} {\PRD}\ }\textbf {\bibinfo {volume} {91}},\ \bibinfo {pages}
  {044009} (\bibinfo {year} {2015})}\BibitemShut {NoStop}%
\bibitem [{\citenamefont {{Zhao}}(2007)}]{Zhao2007.ApJL.671.L1}%
  \BibitemOpen
  \bibfield  {author} {\bibinfo {author} {\bibfnamefont {H.}~\bibnamefont
  {{Zhao}}},\ }\bibinfo {title} {{Coincidences of dark energy with dark matter:
  Clues for a simple alternative?}},\ \href {https://doi.org/10.1086/524731}
  {\bibfield  {journal} {\bibinfo  {journal} {\ApJL}\ }\textbf {\bibinfo
  {volume} {671}},\ \bibinfo {pages} {L1} (\bibinfo {year} {2007})}\BibitemShut
  {NoStop}%
\bibitem [{\citenamefont {Blanchet}\ and\ \citenamefont
  {Le~Tiec}(2008)}]{Blanchet2008.PRD.78.024031}%
  \BibitemOpen
  \bibfield  {author} {\bibinfo {author} {\bibfnamefont {L.}~\bibnamefont
  {Blanchet}}\ and\ \bibinfo {author} {\bibfnamefont {A.}~\bibnamefont
  {Le~Tiec}},\ }\bibinfo {title} {{Model of dark matter and dark energy based
  on gravitational polarization}},\ \href
  {https://doi.org/10.1103/PhysRevD.78.024031} {\bibfield  {journal} {\bibinfo
  {journal} {\PRD}\ }\textbf {\bibinfo {volume} {78}},\ \bibinfo {pages}
  {024031} (\bibinfo {year} {2008})}\BibitemShut {NoStop}%
\bibitem [{Note3()}]{Note3}%
  \BibitemOpen
  \bibinfo {note} {We first fit the \protect \textit {Magneticum} RAR
  result~\cite {Mayer2023.MNRAS.518.257} with a second order polynomial
  $c_0+c_1(1+z)+c_2(1+z)^2$. The result shows $c_1$ is an order of magnitude
  smaller than $c_0$ and $c_2$. Hence the $c_1$ term is negligible. The final
  polynomial $c_0+c_2(1+z)^2$ is just for mathematical
  convenience.}\BibitemShut {Stop}%
\bibitem [{\citenamefont {Ti\'an}(2020)}]{Tian2020.PRD.101.063531}%
  \BibitemOpen
  \bibfield  {author} {\bibinfo {author} {\bibfnamefont {S.~X.}\ \bibnamefont
  {Ti\'an}},\ }\bibinfo {title} {{Cosmological consequences of a scalar field
  with oscillating equation of state: A possible solution to the fine-tuning
  and coincidence problems}},\ \href
  {https://doi.org/10.1103/PhysRevD.101.063531} {\bibfield  {journal} {\bibinfo
   {journal} {\PRD}\ }\textbf {\bibinfo {volume} {101}},\ \bibinfo {pages}
  {063531} (\bibinfo {year} {2020})}\BibitemShut {NoStop}%
\bibitem [{\citenamefont {Peebles}\ and\ \citenamefont
  {Ratra}(1988)}]{Peebles1988.ApJL.325.L17}%
  \BibitemOpen
  \bibfield  {author} {\bibinfo {author} {\bibfnamefont {P.~J.~E.}\
  \bibnamefont {Peebles}}\ and\ \bibinfo {author} {\bibfnamefont
  {B.}~\bibnamefont {Ratra}},\ }\bibinfo {title} {{Cosmology with a
  time-variable cosmological ``constant"}},\ \href
  {https://doi.org/10.1086/185100} {\bibfield  {journal} {\bibinfo  {journal}
  {\ApJL}\ }\textbf {\bibinfo {volume} {325}},\ \bibinfo {pages} {L17}
  (\bibinfo {year} {1988})}\BibitemShut {NoStop}%
\bibitem [{\citenamefont {Steinhardt}\ \emph {et~al.}(1999)\citenamefont
  {Steinhardt}, \citenamefont {Wang},\ and\ \citenamefont
  {Zlatev}}]{Steinhardt1999.PRD.59.123504}%
  \BibitemOpen
  \bibfield  {author} {\bibinfo {author} {\bibfnamefont {P.~J.}\ \bibnamefont
  {Steinhardt}}, \bibinfo {author} {\bibfnamefont {L.}~\bibnamefont {Wang}},\
  and\ \bibinfo {author} {\bibfnamefont {I.}~\bibnamefont {Zlatev}},\ }\bibinfo
  {title} {{Cosmological tracking solutions}},\ \href
  {https://doi.org/10.1103/PhysRevD.59.123504} {\bibfield  {journal} {\bibinfo
  {journal} {\PRD}\ }\textbf {\bibinfo {volume} {59}},\ \bibinfo {pages}
  {123504} (\bibinfo {year} {1999})}\BibitemShut {NoStop}%
\bibitem [{\citenamefont {{Xu}}\ \emph {et~al.}(2022)\citenamefont {{Xu}},
  \citenamefont {{Chen}}, \citenamefont {{Xu}},\ and\ \citenamefont
  {{Cao}}}]{Xu2022.PhysDarkUniverse.36.101023}%
  \BibitemOpen
  \bibfield  {author} {\bibinfo {author} {\bibfnamefont {T.}~\bibnamefont
  {{Xu}}}, \bibinfo {author} {\bibfnamefont {Y.}~\bibnamefont {{Chen}}},
  \bibinfo {author} {\bibfnamefont {L.}~\bibnamefont {{Xu}}},\ and\ \bibinfo
  {author} {\bibfnamefont {S.}~\bibnamefont {{Cao}}},\ }\bibinfo {title}
  {{Comparing the scalar-field dark energy models with recent observations}},\
  \href {https://doi.org/10.1016/j.dark.2022.10102310.48550/arXiv.2109.02453}
  {\bibfield  {journal} {\bibinfo  {journal} {\PhysDarkUniverse}\ }\textbf
  {\bibinfo {volume} {36}},\ \bibinfo {eid} {101023} (\bibinfo {year}
  {2022})}\BibitemShut {NoStop}%
\bibitem [{\citenamefont {Flanagan}\ and\ \citenamefont
  {Hughes}(2005)}]{Flanagan:2005yc}%
  \BibitemOpen
  \bibfield  {author} {\bibinfo {author} {\bibfnamefont {E.~E.}\ \bibnamefont
  {Flanagan}}\ and\ \bibinfo {author} {\bibfnamefont {S.~A.}\ \bibnamefont
  {Hughes}},\ }\bibinfo {title} {{The basics of gravitational wave theory}},\
  \href {https://doi.org/10.1088/1367-2630/7/1/204} {\bibfield  {journal}
  {\bibinfo  {journal} {New J. Phys.}\ }\textbf {\bibinfo {volume} {7}},\
  \bibinfo {pages} {204} (\bibinfo {year} {2005})}\BibitemShut {NoStop}%
\bibitem [{\citenamefont {Gong}\ \emph {et~al.}(2018)\citenamefont {Gong},
  \citenamefont {Hou}, \citenamefont {Liang},\ and\ \citenamefont
  {Papantonopoulos}}]{Gong:2018cgj}%
  \BibitemOpen
  \bibfield  {author} {\bibinfo {author} {\bibfnamefont {Y.}~\bibnamefont
  {Gong}}, \bibinfo {author} {\bibfnamefont {S.}~\bibnamefont {Hou}}, \bibinfo
  {author} {\bibfnamefont {D.}~\bibnamefont {Liang}},\ and\ \bibinfo {author}
  {\bibfnamefont {E.}~\bibnamefont {Papantonopoulos}},\ }\bibinfo {title}
  {{Gravitational waves in Einstein-\ae{}ther and generalized TeVeS theory
  after GW170817}},\ \href {https://doi.org/10.1103/PhysRevD.97.084040}
  {\bibfield  {journal} {\bibinfo  {journal} {Phys. Rev. D}\ }\textbf {\bibinfo
  {volume} {97}},\ \bibinfo {pages} {084040} (\bibinfo {year}
  {2018})}\BibitemShut {NoStop}%
\bibitem [{\citenamefont {Misner}\ \emph {et~al.}(1973)\citenamefont {Misner},
  \citenamefont {Thorne},\ and\ \citenamefont {Wheeler}}]{Misner:1974qy}%
  \BibitemOpen
  \bibfield  {author} {\bibinfo {author} {\bibfnamefont {C.~W.}\ \bibnamefont
  {Misner}}, \bibinfo {author} {\bibfnamefont {K.~S.}\ \bibnamefont {Thorne}},\
  and\ \bibinfo {author} {\bibfnamefont {J.~A.}\ \bibnamefont {Wheeler}},\
  }\href@noop {} {\emph {\bibinfo {title} {{Gravitation}}}}\ (\bibinfo
  {publisher} {W. H. Freeman},\ \bibinfo {address} {San Francisco},\ \bibinfo
  {year} {1973})\BibitemShut {NoStop}%
\bibitem [{\citenamefont {Tian}(2020)}]{Tian2020.PRD.102.063509}%
  \BibitemOpen
  \bibfield  {author} {\bibinfo {author} {\bibfnamefont {S.~X.}\ \bibnamefont
  {Tian}},\ }\bibinfo {title} {{Cosmological consequences of a scalar field
  with oscillating equation of state. II. Oscillating scaling and chaotic
  accelerating solutions}},\ \href
  {https://doi.org/10.1103/PhysRevD.102.063509} {\bibfield  {journal} {\bibinfo
   {journal} {\PRD}\ }\textbf {\bibinfo {volume} {102}},\ \bibinfo {pages}
  {063509} (\bibinfo {year} {2020})}\BibitemShut {NoStop}%
\end{thebibliography}
\end{document}